\documentclass[aps, pra, reprint, showpacs, groupedaddress, superscriptaddress, amsfonts, amsmath, amssymb, nofootinbib]{revtex4-1}

\usepackage{lmodern}
\usepackage{microtype}
\usepackage{bbold}
\usepackage{braket}
\usepackage{array}

\usepackage{hyperref}
\usepackage{graphicx}
\usepackage{bm}
\usepackage{siunitx}
\usepackage{xcolor}
\usepackage{tabularx} 

\definecolor{darkblue}{rgb}{0, 0, 0.8}
\hypersetup{
	pdftitle={Calculation of Rydberg interaction potentials}, 
	colorlinks=true,
	linkcolor=darkblue,
	citecolor=darkblue,
	filecolor=darkblue,
	urlcolor=darkblue
}

\AtBeginDocument{\usepackage{booktabs}}

\usepackage{etoolbox}
\makeatletter
\def\narrowfrontmatter{\relax\rightskip=1.5in\relax}
\patchcmd\section
  {\centering}{\raggedright}{}{Patch failed!}
\patchcmd{\subsection}
  {\centering}{\raggedright}{}{Patch failed!}
\patchcmd\frontmatter@title@format
  {\centering}{\raggedright}{}{Patch failed!}
\patchcmd\frontmatter@authorformat
  {\centering}{\raggedright\narrowfrontmatter}{}{Patch failed!}
\patchcmd\frontmatter@above@affiliation@script
  {\centering}{\raggedright\narrowfrontmatter}{}{Patch failed!}
\patchcmd\frontmatter@RRAP@format
  {\centering}{\raggedright\narrowfrontmatter}{}{Patch failed!}
\apptocmd\adjust@abstractwidth
  {\relax\narrowfrontmatter}{}{Patch failed!}
\renewcommand\frontmatter@abstractwidth{\textwidth}
\makeatother

\begin{document}

\title{Tutorial \\ \fontsize{0.8cm}{1cm}\selectfont Calculation of Rydberg interaction potentials}

\author{Sebastian Weber}
\email{weber@itp3.uni-stuttgart.de}
\affiliation{Institute for Theoretical Physics III and Center for Integrated Quantum Science and Technology, Universit\"{a}t Stuttgart, Pfaffenwaldring 57, 70569 Stuttgart, Germany}

\author{Christoph Tresp}
\affiliation{5. Physikalisches Institut and Center for Integrated Quantum Science and Technology, Universit\"at Stuttgart, Pfaffenwaldring 57, 70569 Stuttgart, Germany}
\affiliation{Department of Physics, Chemistry and Pharmacy, University of Southern Denmark, Campusvej 55, 5230 Odense M, Denmark}

\author{Henri Menke}
\affiliation{Max Planck Institute for Solid State Research, Heisenbergstra\ss e 1, 70569 Stuttgart, Germany}

\author{Alban Urvoy}
\affiliation{5. Physikalisches Institut and Center for Integrated Quantum Science and Technology, Universit\"at Stuttgart, Pfaffenwaldring 57, 70569 Stuttgart, Germany}
\affiliation{Department of Physics and Research Laboratory of Electronics, Massachusetts Institute of Technology, Cambridge, Massachusetts 02139, USA}

\author{Ofer Firstenberg}
\affiliation{Department of Physics of Complex Systems, Weizmann Institute of Science, Rehovot 76100, Israel}

\author{Hans Peter B\"{u}chler}
\affiliation{Institute for Theoretical Physics III and Center for Integrated Quantum Science and Technology, Universit\"{a}t Stuttgart, Pfaffenwaldring 57, 70569 Stuttgart, Germany}

\author{Sebastian Hofferberth}
\email{hofferberth@sdu.dk}
\affiliation{5. Physikalisches Institut and Center for Integrated Quantum Science and Technology, Universit\"at Stuttgart, Pfaffenwaldring 57, 70569 Stuttgart, Germany}
\affiliation{Department of Physics, Chemistry and Pharmacy, University of Southern Denmark, Campusvej 55, 5230 Odense M, Denmark}

\date{\today}

\begin{abstract}
The strong interaction between individual Rydberg atoms provides a powerful tool exploited in an ever-growing range of applications in quantum information science, quantum simulation, and ultracold chemistry. One hallmark of the Rydberg interaction is that both its strength and angular dependence can be fine-tuned with great flexibility by choosing appropriate Rydberg states and applying external electric and magnetic fields. More and more experiments are probing this interaction at short atomic distances or with such high precision that perturbative calculations as well as restrictions to the leading dipole-dipole interaction term are no longer sufficient. In this tutorial, we review all relevant aspects of the full calculation of Rydberg interaction potentials. We discuss the derivation of the interaction Hamiltonian from the electrostatic multipole expansion, numerical and analytical methods for calculating the required electric multipole moments, and the inclusion of electromagnetic fields with arbitrary direction. We focus specifically on symmetry arguments and selection rules, which greatly reduce the size of the Hamiltonian matrix, enabling the direct diagonalization of the Hamiltonian up to higher multipole orders on a desktop computer. Finally, we present example calculations showing the relevance of the full interaction calculation to current experiments. Our software for calculating Rydberg potentials including all features discussed in this tutorial is available as open source.
\end{abstract}

\maketitle

\section{Introduction}
Among the many fascinating properties of highly excited Rydberg atoms \cite{Gallagher1994}, the strong interaction between pairs of Rydberg atoms has proven to be the key feature for diverse applications in quantum information processing and quantum simulation \cite{Saffman2010}. A particularly important concept is the Rydberg blockade \cite{Lukin2000b}, where the excitation of two or more atoms to a Rydberg state is prevented due to the interaction. The Rydberg blockade of atomic ensembles \cite{Lukin2001c} has been observed in ultracold atomic systems in the frozen Rydberg-gas regime \cite{Gallagher1998,Pillet1998}, both in bulk ensembles \cite{Weidemueller2004,Gould2004,Raithel2005,Pillet2006,Pfau2007b,vanLinden2008,Raithel2008,Pfau2008,Pfau2008b,Weidemueller2008,Entin2010,Raithel2011} and in small systems supporting only a single excitation \cite{Kuzmich2012c,Walker2014,Ott2015,Gross2015c}. More recently, experiments have also begun to probe Rydberg interaction effects in room-temperature thermal vapor \cite{Pfau2013b,Adams2013b,Loew2015}. Based on the Rydberg blockade, atomic two-qubit gates have been demonstrated \cite{Saffman2009,Grangier2009,Grangier2010,Saffman2010} as basic building blocks for large-scale neutral atom quantum registers \cite{Saffman2015,Regal2015,Ahn2016,Browaeys2016c,Lukin2016c}. In turn,  the power-law decay of the Rydberg interaction provides interaction over long range and facilitates the extension to multi-qubit Rydberg-mediated gates in such registers \cite{Molmer2008,Zoller2009,Zoller2010b}. Such tailored atomic ensembles are also ideal for investigating processes such as excitation transfer \cite{Gallagher1998,Weidemueller2013b,Browaeys2014,Browaeys2015,Cheinet2016b} or simulation of spin systems \cite{Bloch2015,Browaeys2016b}.

The mapping of Rydberg interactions onto photons by means of electromagnetically induced transparency (EIT) \cite{Hofferberth2016d} has emerged as a powerful approach to realizing few-photon optical nonlinearities \cite{Kurizki2005,Adams2010,Lukin2011,Kurizki2011,Pohl2011b,Vuletic2012,Weidemueller2013}, enabling a variety of optical quantum information applications such as highly efficient single-photon generation \cite{Kuzmich2012b}, entanglement generation between light and atomic excitations \cite{Kuzmich2013}, single-photon all-optical switches \cite{Duerr2014} and transistors \cite{Hofferberth2014,Rempe2014b,Hofferberth2016}, single-photon subtraction \cite{Hofferberth2016e}, and interaction-induced $\pi$-phase shifts \cite{Duerr2016}. Additionally, Rydberg EIT provides access to novel phenomena such as attractive interaction between single photons \cite{Vuletic2013b}, crystallization of photons \cite{Fleischhauer2013}, or photonic scattering resonances \cite{Buechler2014}, as well as spatially resolved detection of single Rydberg atoms in a bulk medium \cite{Lesanovsky2011,Weidemueller2012,Weidemueller2013b}. Recent Rydberg EIT experiments, which simultaneously use Rydberg $S$- and $P$-states \cite{Adams2013} or two different $S$-states \cite{Weidemueller2013b,Hofferberth2014,Rempe2014b} further increase the flexibility of manipulating weak light fields \cite{Lesanovsky2014,Wu2015}.

Detailed understanding of the Rydberg interaction is also required for the concept of Rydberg dressing, where a small Rydberg admixture modifies the interaction between ground-state atoms in ultracold gases \cite{Zoller2010,Pohl2010,Buechler2010,Pupillo2013,Pfau2014b,Pohl2016}. In particular, by the choice of the Rydberg state, one can map the anisotropy of the Rydberg interaction onto the ground-state atoms \cite{Zoller2014c,Zoller2015b,Pohl2015}. Experimental demonstrations of Rydberg dressing have recently been performed using individual atoms \cite{Biedermann2016} or atomic ensembles in an optical lattice \cite{Gross2016}. Finally, the rich structure of the Rydberg interaction potentials supports bound molecular states formed by two Rydberg atoms \cite{Cote2002}, which have been observed in experiments \cite{Gould2003,Shaffer2009,Deiglmayr2016}. Prediction of the equilibrium distance and vibrational spectra for these macro-dimers requires precise knowledge of the interaction potential \cite{Shaffer2006,Gould2008,Deiglmayr2016,Deiglmayr2016b}.

The physics of Rydberg interaction has been well-established for decades \cite{Margenau1939}. As long as the two atoms are well separated and their wave functions do not overlap, one needs to consider only the electrostatic interaction between two localized charge distributions, most conveniently utilizing the well-known electric multipole expansion in spherical coordinates \cite{Rose1958,Fontana1961,Dalgarno1966}. The leading relevant term in this expansion is the dipole-dipole interaction \cite{Ostrovsky2005}, which for unperturbed Rydberg atoms at large separation results in the extensively studied van-der-Waals interaction \cite{Cote2005,Raithel2007,Saffman2008,Pillet2010}. More generally, as long as the interaction energies are small compared to the level spacing of the unperturbed Rydberg pair states, perturbative calculations offer a very convenient method for determining the radial \cite{Cote2005} and angular \cite{Raithel2007,Saffman2008} behavior of the Rydberg potentials.

Nevertheless, the rapid experimental progress in recent years has led to a growing number of experiments for which the perturbative calculation is no longer sufficient. For example, this approach fails when shorter atomic distances are probed and state-mixing due to the interaction becomes significant. The dipole-quadrupole contribution to the interaction has recently been observed both in ultracold \cite{Merkt2014} and room-temperature \cite{Loew2015} systems in experiments with large excitation bandwidth, while the correct prediction of macro-dimer photo-association spectra required the inclusion of terms up to octupolar order in the potential calculation \cite{Deiglmayr2016,Deiglmayr2016b}. In experiments utilizing very high principal quantum numbers $n > 100$~\cite{Pfau2013,Dunning2015,Hofferberth2015}, state-mixing and additional molecular resonances \cite{Lukin2015b} become relevant already at large interatomic distances and make non-perturbative potential calculations necessary.

The nature of the interaction also changes when coupled pair states are (nearly) resonant \cite{Pillet1999,Walker2005}. Such degeneracies, or F\"orster-resonances, can occur naturally or by shifting the pair state energies via external electric~\cite{Gallagher1981,Pillet2006,Raithel2008,Raithel2008b,Entin2010,Pfau2012,Pfau2012b,Weidemueller2013b,Marcassa2016} or microwave~\cite{Martin2004,Martin2007,Tretyakov2014} fields. Such resonances can greatly enhance the Rydberg interaction strength \cite{Browaeys2015b}. When spin-orbit coupling as well as Stark and Zeeman splitting of all involved levels are taken into account, the resonances can exhibit new features and a rich angular dependence \cite{Hofferberth2016,Hofferberth2016f}. Full potential calculations including the external fields reveal the number of states which must be included in specific cases for accurate results.

As consequence of the rapid evolution of the field, it becomes more and more common to rely on numerical diagonalization of the Rydberg interaction Hamiltonian including higher orders of the multipole expansion \cite{Shaffer2006,Merkt2014,Hofferberth2015,Deiglmayr2016}. One example is the recently released ARC library \cite{Sibalic2016}, which provides powerful tools for calculating alkali Rydberg properties. In this tutorial, we discuss all relevant steps required for the numerical calculation of pairwise Rydberg interaction potentials:
\begin{enumerate}
  \item Construction of the single-atom Hamiltonian from the orbital wave functions (\ref{sec:potential}) and their transition matrix elements (\ref{sec:radialmatrixelements} and \ref{sec:angularmatrixelements}) in the absence of external fields.
  \item Derivation of the interaction Hamiltonian for a pair of Rydberg atoms by multipole expansion (Sec. \ref{sec:hamiltonian}).
  \item Inclusion of external electric and magnetic fields in arbitrary directions relative to the inter-atomic axis (Sec. \ref{sec:externalfields}).
  \item Application of selection rules and symmetry arguments to reduce the size of the Hilbert space to the relevant states (Sec. \ref{sec:conservations}).
  \item Rotation of the interaction Hamiltonian to a particular coordinate system, given for example by the direction of an incident excitation laser (Sec. \ref{sec:angulardependency}).
  \item Diagonalization and extraction of the interaction potentials (Sec. \ref{sec:diagonalization}). In the context of two practical applications (Sec. \ref{sec:Applications}), we discuss best practices and specific considerations in the relevant scenarios. These examples illustrate the capabilities of the presented approach and the agreement with experiments.
\end{enumerate}

Our software for numerical Rydberg potential calculations, which includes all features we discuss in this review, is available open source from \mbox{\url{https://pairinteraction.github.io/}}.

\section{Rydberg interaction}

\subsection{Hamiltonian}
\label{sec:hamiltonian}
\begin{figure*}
	\includegraphics{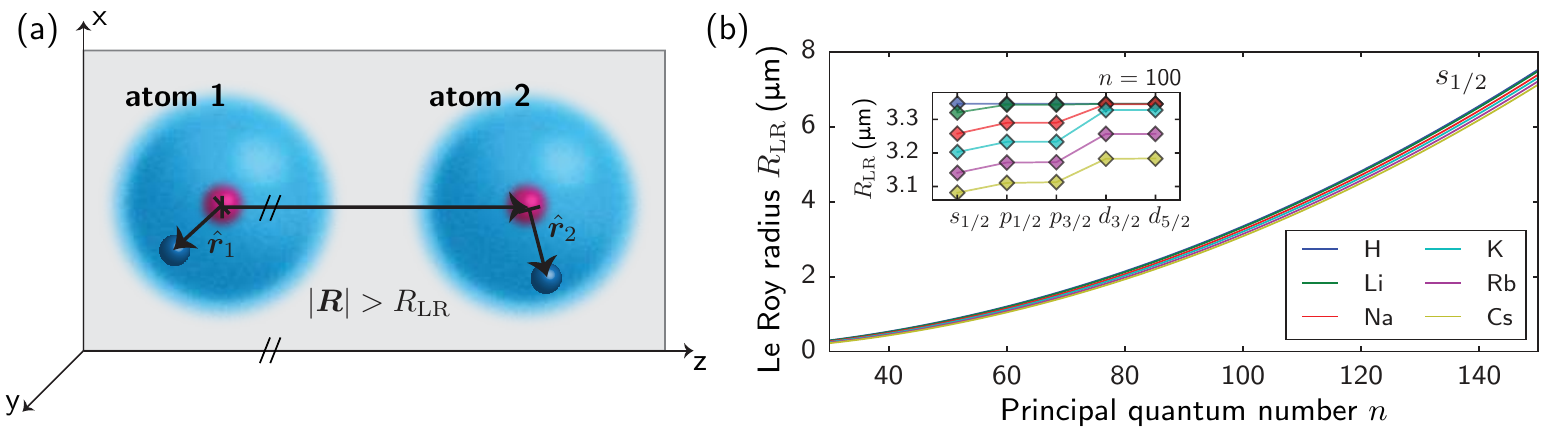}
	\caption{\textsf{(a)} Considered system. We study two Rydberg atoms whose interatomic axis is parallel to the $z$-axis. The positions of the Rydberg electrons are $\hat{r}_1$ and $\hat{r}_2$. The interatomic distance $R$ is larger than the Le~Roy radius $R_\text{LR}$ \cite{LeRoy1974} so that the electronic wave functions do not overlap. \textsf{(b)} Le~Roy radius for pairs of alkali atoms. For calculating the Le~Roy radius, we assumed both atoms to be in the same state. The Le~Roy radius increases approximately with the square of the principal quantum number. The inset shows its dependence on the momentum quantum numbers. The Le~Roy radius for alkali atoms is bounded from above by the Le~Roy radius for hydrogen atoms $R_\text{LR}=a_0 \sqrt{8 n^2 (5n^2+1-3l(l+1))}$ \cite{Bethe1957} where $a_0$ is the Bohr radius.}
        \label{fig:system}
\end{figure*}

We study two neutral atoms, each having one electron excited into a Rydberg state, as depicted in Figure \ref{fig:system} \textsf{(a)}. Because we are only interested in interatomic distances $R$ for which the Rydberg atoms are well-separated, the interaction between the atoms is dominated by the strong interactions of the
Rydberg electrons. Using the Born-Oppenheimer approximation \cite{Born1927}, the corresponding two-atom Hamiltonian is of the form
\begin{equation}
\hat{H}(\bm{R}) = \hat{H}_0 + \hat{H}_{\text{int}}(\bm{R}) \; ,
\label{eqn:hamiltonian}
\end{equation}
where $\hat{H}_0$ contains the energies of the unperturbed Rydberg states. The operator $\hat{H}_{\text{int}}$ captures the interaction between the two Rydberg electrons, the two ionic cores, and the Rydberg electron of one atom and the ionic core of the other atom. This is the standard treatment of two interacting Rydberg atoms which is discussed in similar detail in \cite{Ostrovsky2005, Cote2005, Shaffer2009,Deiglmayr2016b}.

Because the hyperfine splitting of Rydberg levels is much smaller than typical interaction energies \cite{Merkt2013,Gallagher2003,Spreeuw2013}, we use the fine-structure basis. As we will see later, we can assume the two Rydberg electrons to be distinguishable particles. Hence, we use the product basis  $ \ket{n_1l_1j_1 m_{j1};n_2l_2j_2
m_{j2}}=\ket{n_1l_1j_1 m_{j1}} \otimes \ket{n_2l_2j_2 m_{j2}}$\footnote{We omit the spin quantum number $s_1 = s_2 =1/2$ in our notation.}. Note that the two Rydberg atoms are allowed to be of different chemical species \cite{Beterov2015, Zeng2017}. The operator $\hat{H}_0$ can be written as
\begin{align}
\hat{H}_0 = \sum\limits_{n_1,l_1,j_1,m_{j1}} E_{n_1l_1j_1} \ket{n_1l_1j_1m_{j1}}\bra{n_1l_1j_1m_{j1}} \otimes \mathbb{1} \nonumber\\
+ \mathbb{1} \otimes \sum\limits_{n_2,l_2,j_2,m_{j2}} E_{n_2l_2j_2} \ket{n_2l_2j_2m_{j2}}\bra{n_2l_2j_2m_{j2}} \; .
\end{align}
We use the common convention that the quantization axis points along the $z$-direction. The potential energy $E_{nlj}$ of an electron excited to a Rydberg state is given by a formula\footnote{Note that we use SI units throughout the manuscript.} similar to the one known for the hydrogen atom
\begin{equation}
E_{nlj} = - \frac{h c R^*}{(n-\delta_{nlj})^2} \; ,
\label{eq:RydberglevelEnergy}
\end{equation}
where $R^*$ is the modified Rydberg constant and $\delta_{nlj}$ is the quantum defect \cite{Seaton1983}. These species-dependent parameters are used to capture subtle differences between the bare Coulomb potential of a hydrogen core and the actual potential felt by the Rydberg electron. For details on these parameters, see \ref{sec:potential}.

For calculating the interaction energy $\hat{H}_{\text{int}}$, we neglect retardation effects \cite{Casimir1948} as the wavelengths of the involved Rydberg-Rydberg transitions are much larger than the considered interatomic distances. Furthermore, we assume the interatomic distance to be larger than the Le~Roy radius \cite{LeRoy1974}
\begin{equation}
R_{\text{LR}} = 2 \left( \sqrt{\braket{n_1l_1j_1 | \hat{r}^2|  n_1l_1j_1}} + \sqrt{\braket{n_2l_2j_2 | \hat{r}^2|  n_2l_2j_2 }} \right) \; ,
\end{equation}
which increases approximately with the square of the principal quantum number like the radius of a Rydberg atom does, see Figure \ref{fig:system} \textsf{(b)}. This assumption tremendously simplifies the calculations. It ensures that the electronic wave functions do not overlap, so that exchange interaction and charge overlap interaction can be neglected. Thus, we can treat the electrons as distinguishable particles.  Furthermore, it allows us to use a multipole expansion for the interaction energy. In order to do so, we first think of the two Rydberg atoms as classical charge distributions \cite{Jackson1998}. Their electrostatic interaction energy is given by
\begin{align}
H_{\text{int}}(\bm{R})
=&
\frac{e^2}{4 \pi \epsilon_0} \bigg(
\frac{1}{ |\bm{R} + \bm{r}_2 - \bm{r}_1|}
+  \frac{1}{|\bm{R}|} \nonumber\\
&- \frac{1}{|\bm{R} - \bm{r}_1|}
- \frac{1}{ |\bm{R} + \bm{r}_2|}
\bigg)
\; ,
\end{align}
where $\bm{R}$ denotes the distance vector between the atoms. The positions $\bm{r}_1$ and $\bm{r}_2$ of the electrons of the first and second atom are given as relative coordinates in the body frame of the respective atom, see Figure \ref{fig:system} \textsf{(a)}. The multipole expansion leads to \cite{Rose1958, Fontana1961, Dalgarno1966}
\begin{equation} \label{eqn:MultiPoleExpansion}
H_{\text{int}}(\bm{R}) = \sum_{\kappa_1,\kappa_2=1}^\infty \frac{V_{\kappa_1\kappa_2}}{4 \pi \epsilon_0 |\bm{R}|^{\kappa_1 + \kappa_2 + 1}} \; .
\end{equation}
The exact form of $V_{\kappa_1\kappa_2}$ depends on the choice of the coordinate systems used to label the positions of the electrons. If we choose the coordinate systems such that the $z$-axis points along $\bm{R}$, i.e. along the interatomic axis, we get the comparatively simple result
\begin{equation}
V_{\kappa_1\kappa_2} = (-1)^{\kappa_2} \mkern-10mu \sum_{q=-{\kappa_<}}^{\kappa_<} \sqrt{
	\binom{\kappa_1 + \kappa_2}{\kappa_1 + q}
	\binom{\kappa_1 + \kappa_2}{\kappa_2 + q}
} \, p_{\kappa_1 q}^{(1)}  p_{\kappa_2 -q}^{(2)},
\label{eqn:multipole}
\end{equation}
where we use $\kappa_< = \min(\kappa_1, \kappa_2)$ and binomial coefficients to shorten our notation. This result transfers into quantum mechanics by canonical quantization. Thus, the spherical multipole moments $p_{\kappa q}^{(1)}$ and $p_{\kappa q}^{(2)}$ become the spherical multipole operators $\hat{p}_{\kappa q}^{(1)}$ and $\hat{p}_{\kappa q}^{(2)}$, that operate on the Rydberg electron of the first and second atom, respectively. The operators are of the form
\begin{equation}
\hat{p}_{\kappa q}^{(i)} =  e \, \hat{r}^{\kappa}_i   \cdot \sqrt{\frac{4 \pi}{2\kappa+1}} Y_{\kappa  q}(\hat{\vartheta}_i ,\hat{\varphi}_i )\;, \label{eqn:sphericalmultipole}
\end{equation}
where $Y_{ \kappa q}(\hat{\vartheta},\hat{\varphi})$ are spherical harmonics\footnote{In the literature, different normalizations for spherical harmonics are found. We choose the convention $Y_{\kappa q}(\vartheta,\varphi) =  (-1)^q \sqrt{\frac{(2\kappa+1)(\kappa-q)!}{4 \pi (\kappa + q)!}} \sin^q \vartheta  \frac{\mathrm{d}^q}{(\mathrm{d}\cos \vartheta)^q} P_\kappa(\cos \vartheta)\;\text{e}^{\mathrm{i}q\varphi}$ that is commonly used in quantum mechanics. Here,  $P_\kappa$ are Legendre polynomials and $(-1)^q$ the Condon-Shortley phase.}. Note that, in our notation, the spherical basis is $\{ \bm{e}_\pm = \mp \frac{1}{\sqrt{2}} (\bm{e}_x \mp i \bm{e}_y),~\bm{e}_0 =\bm{e}_z \}$\footnote{We use the common definition of the spherical multipole operators~(\ref{eqn:sphericalmultipole}). This implies that our spherical basis is non-standard (with the standard convention, the dipole operator in the spherical basis would not be of the usual form $\hat{p}_{1 1} \bm{e}_+ + \hat{p}_{1 -1} \bm{e}_- + \hat{p}_{1 0} \bm{e}_0$).}. The spherical multipole operator $\hat{p}_{\kappa q}$ corresponds to the $2^\kappa$-pole momentum. The multipole expansion (\ref{eqn:MultiPoleExpansion}) is a series expansion of the interaction potential in powers
\begin{equation}
\label{eqn:multipoleorder}
\varrho = \kappa_1 + \kappa_2 + 1\;
\end{equation}
of the inverse interatomic distance. The series expansion starts at $\varrho = 3$. Thus, the contribution of lowest order is the dipole-dipole interaction which reflects the neutral charge of the Rydberg atoms. Section \ref{sec:higherOrders} discusses the relevance of higher-order contributions. In general, the order at which we can reasonably truncate the expansion increases with decreasing interatomic distance.

The spherical multipole operators are composed of the product of a radial and an angular operator whose matrix elements can be calculated independently from one another with the formalism shown in \ref{sec:radial} and \ref{sec:angularmatrixelements}. The independent calculation works because the potential for the Rydberg electron is spherically symmetric, so that the Rydberg wave function can be separated into the product of a radial function $\Psi^{\text{rad}}_{nlj}(r)$ and a spin spherical harmonic $Y_{lsjm_j}(\vartheta,\varphi) $ \cite{Biedenharn2009},

\begin{equation}
\Psi(r,\vartheta,\varphi) = \Psi^{\text{rad}}_{nlj}(r) \cdot Y_{lsjm_j}(\vartheta,\varphi) \;. \label{eqn:seperation}
\end{equation}

\subsection{External fields}\label{sec:externalfields}

\begin{figure*}
	\includegraphics{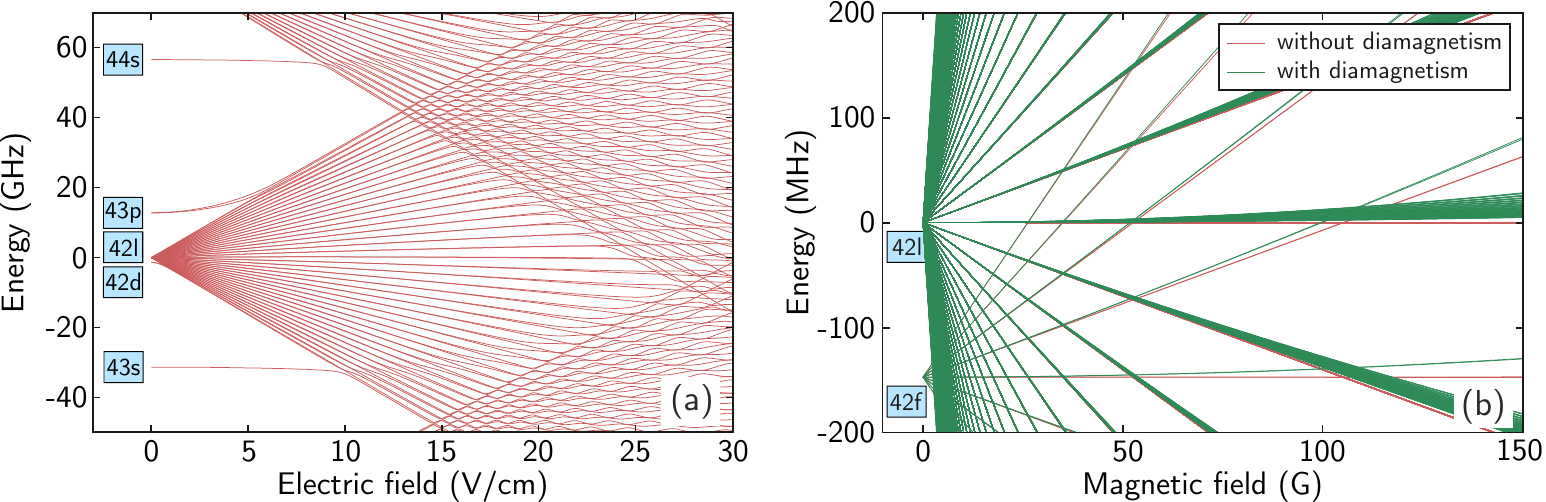}
	\caption{\textsf{(a)} Stark map for the Na atom in the energy range of the $n=42$ manifold. The magnetic quantum numbers do not mix under the assumption that the quantization axis is chosen parallel to the field. For clarity, only states with $m=1/2$ are shown. \textsf{(b)} Zeeman map for the Na $n=42$ manifold. Here, all the states within the plot range are depicted. The diamagnetic interaction increases the energies of the states. However, as diamagnetism does not contribute to the linear Zeeman effect it can be neglected for a weak magnetic field.} \label{fig:StarkZeeman}
\end{figure*}

In general, the interaction between an atom and the electromagnetic field can be taken care of by employing the minimal-coupling replacement \cite{Steck2007, Milonni2013}. Assuming that the fields are static and homogeneous, this general approach gets reduced to adding the electric interaction
\begin{equation}
\hat{V}_{\text{e}} = -\hat{\bm{d}} \cdot \bm{E} \quad\text{with}\quad \hat{\bm{d}} = e \hat{\bm{r}} \label{eqn:electric}
\end{equation}
and the magnetic interaction
\begin{equation}
\hat{V}_{\text{m}} = -\hat{\bm{\mu}} \cdot \bm{B} + \frac{1}{8m_e} | \hat{\bm{d}} \times \bm{B}|^2 \quad\text{with}\quad \hat{\bm{\mu}} = - \frac{\mu_{\text{B}}}{\hbar} (g_l \hat{\bm{l}}+g_s \hat{\bm{s}}) \label{eqn:magnetic}
\end{equation}
to the Hamiltonian (\ref{eqn:hamiltonian}), where $\hat{\bm{d}}$ is the electric dipole operator, and $\hat{\bm{\mu}}$ is the magnetic dipole operator. The $g$-factors $g_l$ and $g_s$ characterize the magnetic moment through orbital motion and spin. The constant $\mu_{\text{B}}$ is the Bohr magneton. The term $\frac{1}{8m_e} | \hat{\bm{d}} \times \bm{B}|^2$ is the diamagnetic interaction. Most importantly, these formulas allow for electromagnetic fields in arbitrary directions and in particular facilitate arbitrary angles between magnetic and electric fields. The final Hamiltonian is
\begin{align}
\hat{H}(\bm{R}) =& \hat{H}_0 + \hat{H}_{\text{int}}(\bm{R}) + \hat{V}_{\text{e}}  \otimes \mathbb{1}  + \mathbb{1} \otimes \hat{V}_{\text{e}}  \nonumber\\
&+ \hat{V}_{\text{m}} \otimes \mathbb{1} + \mathbb{1} \otimes  \hat{V}_{\text{m}} \; .
\label{eqn:hamiltoniantotal}
\end{align}

In order to calculate matrix elements of $\hat{V}_{\text{e}}$ and $\hat{V}_{\text{m}}$, we have to do some preparatory work. In \ref{sec:angularmatrixelements}, we review a powerful formalism to calculate matrix elements of spherical tensor operators. To make use of it, the notation of equations (\ref{eqn:electric}) and (\ref{eqn:magnetic}) has to be changed from the cartesian basis to the spherical basis  $\{ \bm{e}_\pm = \mp \frac{1}{\sqrt{2}} (\bm{e}_x \mp i \bm{e}_y),~\bm{e}_0 =\bm{e}_z \}$. In the spherical basis the components of, for example, the electric field are given by
\begin{equation}
{E}_\pm = \mp \frac{1}{\sqrt{2}} ({E}_x \pm i {E}_y)\;,~{E}_0 ={E}_z\;.
\end{equation}
Using this expression, the interaction with an electric field reads
\begin{equation}
-\hat{\bm{d}} \cdot \bm{E} = -e\,\hat{r} \cdot \sqrt{\frac{4\pi}{3}}\left(\hat{Y}_{1,0} {E}_0 - \hat{Y}_{1,1} {E}_- -  \hat{Y}_{1,-1} {E}_+ \right)\;
\end{equation}
with spherical harmonics $Y_{ \kappa q}(\hat{\vartheta},\hat{\varphi})$. Likewise we can express the dot product of $\hat{\bm{J}} \in \{\hat{\bm{l}}, \hat{\bm{s}}\}$ and the magnetic field in the spherical basis
\begin{equation}
\hat{\bm{J}} \cdot \bm{B} = \hat{J}_{1,0} {B}_0 - \hat{J}_{1,1} {B}_- - \hat{J}_{1,-1} {B}_+\;,
\end{equation}
where $\hat{J}_{1q}$ are the spherical momentum operators. The operators $\hat{J}_{1,\pm1}$ are related to the ladder operators $\hat{J}_{\pm} = \mp \sqrt{2}  \hat{J}_{1,\pm1} $ and $\hat{J}_{z} =\hat{J}_{1,0}$. The diamagnetic interaction in the spherical basis reads
\begin{align}
\frac{1}{8m_e} |\hat{\bm{d}} \times &  \bm{B}|^2  = \frac{e^2}{12m_e} \hat{r}^2 \cdot \sqrt{\frac{4\pi}{5}}\biggl(
\sqrt{5} \hat{Y}_{0,0} {B}^2 \nonumber\\
&- \hat{Y}_{2,0} ( {B}_0{B}_0 + {B}_+{B}_- ) \nonumber\\
&+ \sqrt{3} \hat{Y}_{2,1}  {B}_0{B}_- + \sqrt{3} \hat{Y}_{2,-1}  {B}_0{B}_+ \nonumber\\
&- \sqrt{\frac{3}{2}} \hat{Y}_{2,2} {B}_-{B}_- - \sqrt{\frac{3}{2}} \hat{Y}_{2,-2} {B}_+{B}_+ \biggr)\;. \label{sec:diamagnetic}
\end{align}

The entire atom-field interaction is now expressed in terms of spherical tensor operators, and we can proceed to calculate the matrix elements via the formalism of \ref{sec:radial} and \ref{sec:angularmatrixelements}. This enables not only the calculation of pair potentials in the presence of external fields but also the computation of Stark/Zeeman maps, see Figure \ref{fig:StarkZeeman}.

\subsection{Selection rules and symmetries}
\label{sec:conservations}

Knowing how to calculate matrix elements facilitates a rigorous derivation of the selection rules. The results for spherical harmonics and momentum operators are shown in Table~\ref{tab:selectionrules}. Note that the selection rules for spherical harmonics directly apply to the multipole operators. The selection rules greatly reduce the number of matrix elements which must be calculated explicitly for the construction of the pair Hamiltonian, enabling a significant reduction of computation time.

\begin{table}[!ht]
	\caption{\label{tab:selectionrules}Selection rules for matrix elements of spherical harmonics $\braket{nlsjm_j | \hat{Y}_{ \kappa q} | n'l's'j'm_j'}$ and momentum operators  $\braket{lsjm_j | \hat{J}_{1q} | l's'j'm_j'}$ where $\hat{J}_{1q} \in \{\hat{l}_{1q}, \hat{s}_{1q}\}$. The selection rules for spherical harmonics equal the selection rules for matrix elements of spherical multipole operators $\braket{nlsjm_j | \hat{p}_{ \kappa q} | n'l's'j'm_j'}$. The stated selection rules are explicitly tailored towards our use-case of atoms with one single Rydberg electron. They do not hold true for multi-electron atoms or molecules.}
		\begin{tabularx}{\linewidth}{Xl}
			\toprule
			Spherical harmonics $\hat{Y}_{\kappa q}$ and  & Momentum operators \\
			multipole operators $\hat{p}_{\kappa q}$  & $\hat{J}_{1q}$\\
			\midrule 
			$n$ not restricted&$n$ not restricted\\
			$l= \begin{cases} l' \pm 0, 2, ..., \kappa & \text{for even $\kappa$} \\ l' \pm 1, 3, ..., \kappa & \text{for odd $\kappa$} \end{cases} $ & $l=l'$ \\
			$s=s'$ & $s=s'$ \\
			$j=j' \pm 0, 1, ..., \kappa$  and $j+j'\geq \kappa$  &  $j=j' \pm 0, 1$\\
			$m_j=m_j'+q$ & $m_j=m_j'+q$ \\
			\hspace{1em} with $q \in \{-\kappa,-\kappa+1, ..., \kappa\}$ & \hspace{1em} with $q \in \{-1,0,1\}$ \\
			\bottomrule
		\end{tabularx}
\end{table}

\begin{table*}
	\caption{\label{tab:symmetries}
		Overview of which symmetry operation commutes with which part of the Hamiltonian (\ref{eqn:hamiltoniantotal}). A $\checkmark$-sign indicates that the symmetry is conserved whereas a $\diagup$-sign signals that the symmetry is broken. In case of the atom-field interactions $\hat{V}_m$ and $\hat{V}_e$, we distinguish between fields in the $x$, $y$, and $z$-direction. Inversion and permutation symmetry are only present in homonuclear systems. The reflection symmetry is only of importance if $m_{j1}+m_{j2}=0$. Note, if both inversion and permutation symmetry are present, $(-1)^{l_1+l_2}$ is conserved as well. }
		\begin{tabularx}{\linewidth}{@{\quad}p{14em}p{6em}p{6em}p{9em}XXXXXX}
			\toprule
			Symmetry  & Conserved & \multicolumn{2}{l}{$\hat{H}_\text{int}$} & \multicolumn{3}{l}{$\hat{V}_\text{m}$} & \multicolumn{3}{l}{$\hat{V}_\text{e}$} \\ \cmidrule(r){3-4} \cmidrule(r){5-7} \cmidrule(r){8-10}
			operation &  quantity & \scriptsize Dipole-dipole & \scriptsize Up to higher orders & \scriptsize $x$ & \scriptsize $y$ & \scriptsize $z$ & \scriptsize $x$ & \scriptsize $y$ & \scriptsize $z$ \\
			\midrule
			\multicolumn{9}{l}{\textbf{Symmetries originating from the point group $\bm{D_{\infty h}}$}} \\
			Rotation about $z$-axis& $m_{j1}+m_{j2}$ &$\checkmark$  & $\checkmark$  & $\diagup$  & $\diagup$ & $\checkmark$  & $\diagup$  & $\diagup$  & $\checkmark$ \\
			Reflection through $xz$-plane & $+/-$ & $\checkmark$ & $\checkmark$ & $\diagup$ & $\checkmark$ & $\diagup$ & $\checkmark$ & $\diagup$  & $\checkmark$  \\
			Inversion & $g/u$ &$\checkmark$  & $\checkmark$ & $\checkmark$ & $\checkmark$ & $\checkmark$ & $\diagup$ & $\diagup$ & $\diagup$ \\
			\multicolumn{9}{l}{\textbf{Symmetry in case of symmetric interaction potentials}} \\
			Permutation& $s/a$  &$\checkmark$  & $\diagup$ & $\checkmark$ & $\checkmark$ & $\checkmark$ & $\checkmark$& $\checkmark$ & $\checkmark$  \\
			\bottomrule
		\end{tabularx}
\end{table*}

The considered system of two interacting Rydberg atoms has the same symmetries as any diatomic molecule. The point group of the system is $C_{\infty v}$/$D_{\infty h}$ if the system is heteronuclear/homonuclear \cite{Herzberg1950}. The symmetries of the point group are conserved by the Hamiltonian (\ref{eqn:hamiltonian}) of the two Rydberg atoms in the absence of external fields.

From the symmetry under rotation about the interatomic axis, it follows that the projection of the total angular momentum on the interatomic axis is conserved. Since we have chosen the quantization axis to be parallel to the interatomic axis, the total magnetic quantum number
\begin{equation}
M = m_{j1}+m_{j2}
\end{equation}
is conserved.

If the system is homonuclear, a further symmetry is the inversion symmetry ($\bm{r}_i \rightarrow -\bm{r}_i$, $\bm{R}\rightarrow -\bm{R}$).  A properly symmetrized basis state is of the form~\cite{Brown2003, Gould2006, Deiglmayr2016b}
\begin{align}
\ket{\Psi}_{g/u} \propto& \ket{n_1l_1j_1 m_{j1};n_2l_2j_2 m_{j2}}  \nonumber\\
&- p (-1)^{l_1+l_2} \ket{n_2l_2j_2 m_{j2}; n_1l_1j_1 m_{j1}} \;,
\label{eqn:gu}
\end{align}
where $p=+1$ for states with \textit{gerade} symmetry $\ket{\Psi}_g$ and $p=-1$ for states with \textit{ungerade} symmetry $\ket{\Psi}_u$. The Hamiltonian does not couple states of gerade symmetry to states of ungerade symmetry.

Independent of whether the system is homonuclear or heteronuclear, it is symmetric under reflection through a plane containing the interatomic axis~\cite{Brown2003, Gould2006}. We choose the $xz$-plane as mirror plane. The reflection symmetry ($y_i \rightarrow -y_i$) can be exploited by changing into the symmetrized basis
\begin{align}
\ket{\Psi}_{+/-} \propto& \ket{n_1l_1j_1m_{j1};n_2l_2j_2m_{j2}} \nonumber  \\
& +d(-1)^{l_1+l_2+m_{j1}+m_{j2}-j_1-j_2}  \nonumber\\
&\times \ket{n_1l_1j_1-m_{j1};n_2l_2j_2-m_{j2}} \; \label{eqn:pm}
\end{align}
with $d=+1$ for even states $\ket{\Psi}_+$ and $d=- 1$ for odd states $\ket{\Psi}_-$ under reflection. If the total magnetic quantum number is zero, we can symmetrize with respect to rotation, reflection, and inversion simultaneously. If it is non-zero, this is not possible since then the reflection and the rotation do not commute. In this case, we neglect the reflection symmetry.

In case of pure dipole-dipole interaction, a system of two homonuclear Rydberg atoms is subject to permutation symmetry  ($\bm{R}\rightarrow -\bm{R}$) in addition to the symmetries of the point group. The interaction potential is symmetric under exchange of the two ionic cores. A properly symmetrized basis state is
\begin{align}
\ket{\Psi}_{s/a} \propto& \ket{n_1l_1j_1m_{j1};n_2l_2j_2m_{j2}} \nonumber\\
& - f \ket{n_2l_2j_2 m_{j2}; n_1l_1j_1 m_{j1}} \;, \label{eqn:symasym}
\end{align}
where $f=+1$ for \textit{symmetric} states $\ket{\Psi}_s$ and $f=- 1$ for \textit{antisymmetric} states $\ket{\Psi}_a$. Comparing equation (\ref{eqn:gu}) and (\ref{eqn:symasym}) shows that
\begin{equation}
P = (-1)^{l_1+l_2}
\end{equation}
is conserved when both permutation and inversion symmetry are present.

External fields might break the symmetries discussed above. If the operator of the atom-field interaction $\hat{V}_{\text{e}}$ (\ref{eqn:electric}) or $\hat{V}_{\text{m}}$ (\ref{eqn:magnetic}) does not commute with a symmetry operation, the symmetry is not conserved. For example, external fields that do not point along the interatomic axis can mix states of different total magnetic quantum numbers. Table \ref{tab:symmetries} contains an overview of which symmetry operation commutes with which part of the Hamiltonian  (\ref{eqn:hamiltoniantotal}) of two interacting Rydberg atoms in the presence of external fields.

\subsection{Angular dependency}\label{sec:angulardependency}

\begin{figure*}
	\includegraphics{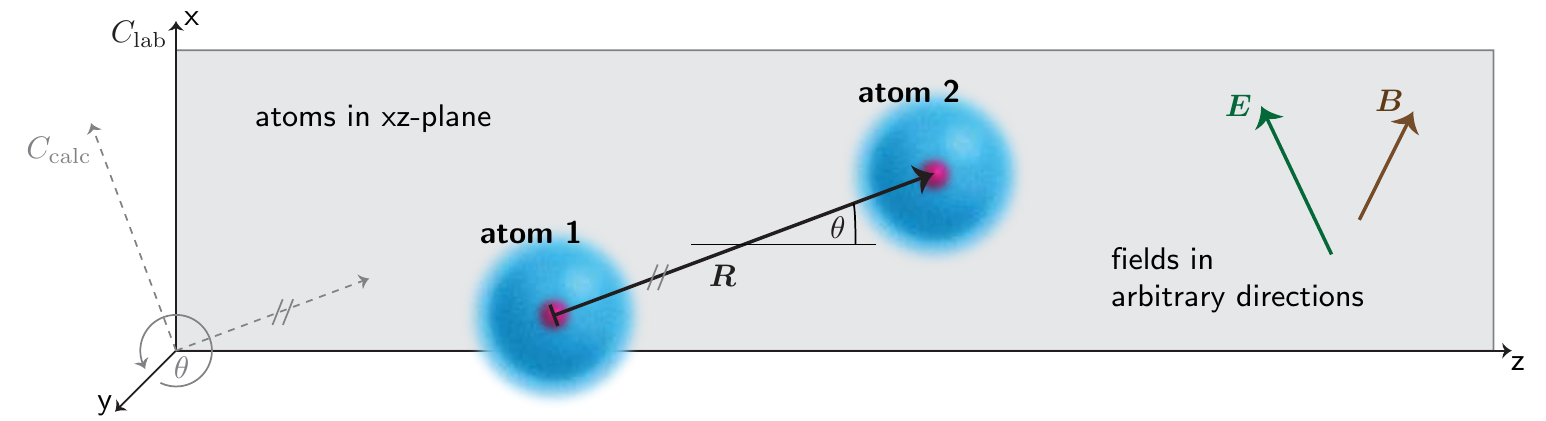}
	\caption{Rotation of the coordinate system. We consider a pair of atoms whose interatomic axis is not parallel to the $z$-axis. In order to apply equation (\ref{eqn:multipole}) for the atom-atom interaction, we have to rotate the coordinate system. Assuming that the interatomic axis lies in the $xz$-plane, a rotation around the $y$-axis is needed. The rotation changes the representation of the quantum mechanical states as well as the representation of the electric and magnetic fields. The figure illustrates our definition of the rotation angle $\theta$.}\label{fig:systemxz}
\end{figure*}

As discussed in section \ref{sec:hamiltonian}, the formula for the atom-atom interaction (\ref{eqn:multipole}) is only valid if we are in a coordinate system  $C_{\text{calc}}$ where the $z$-axis, which we chose as the quantization axis, points in the same direction as the interatomic axis. However, sometimes one would like to use a different coordinate system. For example, when the Rydberg state is excited by a laser pulse, it is convenient to use a coordinate system $C_{\text{lab}}$ where the $z$-axis points along the laser beam.

To ease the calculation, let us assume that the interatomic axis lies in the $xz$-plane of $C_{\text{lab}}$.  Then we can change into $C_{\text{calc}}$ by rotating $C_{\text{lab}}$ around the $y$-axis, see Figure \ref{fig:systemxz}. The rotation angle $\theta$ is defined as the angle between the interatomic axis and the $z$-axis, and is called interaction angle.

Given that our coordinate systems are right-handed and the rotation of the coordinate system is counter-clockwise if the $y$-axis points towards the observer, the magnetic and electric fields transform according to
\begin{equation}
\bm{E}_{\text{calc}} = \left(\begin{array}{@{}lll@{}}
\cos \theta & 0 & -\sin \theta  \\
0 & 1 & 0 \\
\sin \theta & 0 & \cos \theta  \\
\end{array}\right)  \bm{E}_{\text{lab}} \;. \label{eqn:trafo_fields}
\end{equation}

Any state $\ket{nljm_j}_{\text{lab}}$ in the lab frame, i.e.\ with the quantization axis parallel to the laser beam, should be expressed as a linear superposition of states $\ket{nljm_j}_{\text{calc}}$ whose quantization axis points along the $z$-axis of $C_{\text{calc}}$,
\begin{equation}
\ket{nljm_j}_{\text{lab}} = \sum_{m_j'} d_{m_jm_j'}^j(\theta)\ket{nljm_j'}_{\text{calc}} \;,  \label{eqn:trafo_states}
\end{equation}
where the coefficients $d_{m_jm_j'}^j(\theta)$ are elements of the Wigner (lowercase) $d$-matrix \cite{Wigner1959}. If we had not fixed the interatomic axes in the $xz$-plane, we would have to use the Wigner (uppercase) $D$-matrix instead. Our notation is consistent with the definitions in \cite{Varshalovich1988}.

Using the transformations (\ref{eqn:trafo_fields}) and (\ref{eqn:trafo_states}), we can switch to the coordinate system $C_{\text{calc}}$ and perform all calculations there.

In the absence of external electromagnetic fields, the eigenvalues of the Hamiltonian and thus the pair potentials do not depend on the interaction angle. However, the \emph{overlap} of each unperturbed state $\ket{n_1l_1j_1 m_{j1}; n_2l_2j_2 m_{j2}}$ with a particular eigenstate does depend on the interaction angle. The quantum numbers of the unperturbed state become good quantum numbers if the interatomic distance approaches infinity and the atoms are not perturbed by external fields.

\begin{figure*}
	\includegraphics{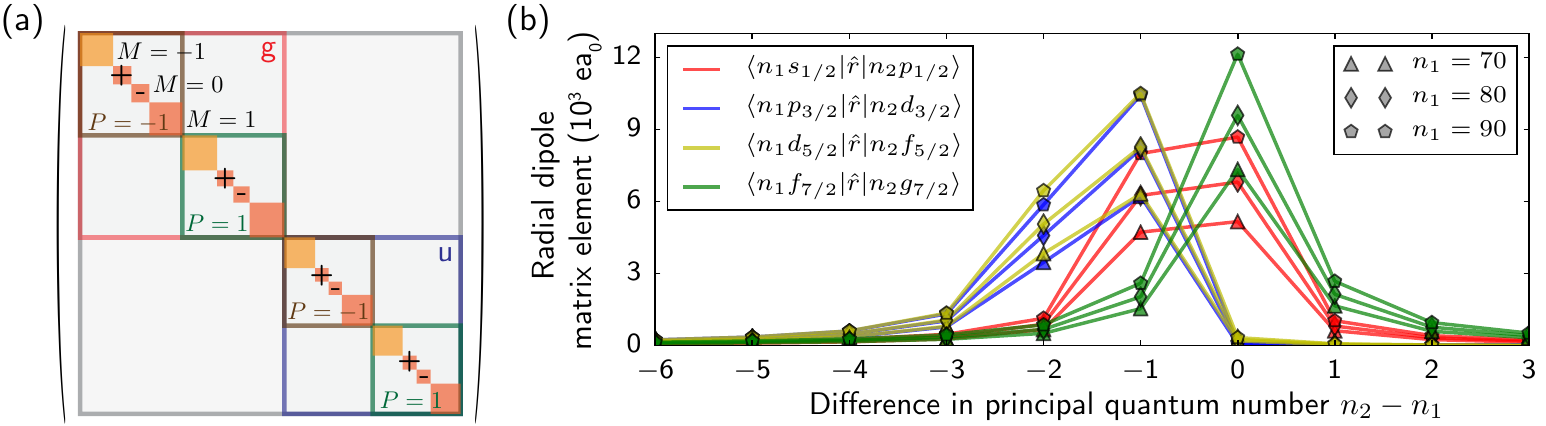}
	\caption{\textsf{(a)} Block diagonality of the Hamiltonian matrix for a pair of dipole-dipole interacting Rydberg atoms in the absence of external fields. As discussed in section \ref{sec:conservations}, the Hamiltonian matrix for a homonuclear pair of atoms decomposes into a gerade/ungerade block (g/u) . Pure dipole-dipole interaction additionally conserves $P=(-1)^{l_1+l_2}$. Moreover, $M=m_{j1}+m_{j2}$ is conserved if the interatomic axis points along the $z$-axis. Blocks belonging to $M=0$ can be further decomposed into an even/odd block under reflection (+/-). \textsf{(b)} Radial dipole matrix elements for rubidium as a function of the difference in principal quantum numbers $n_2-n_1$ (for reasons of simplicity, only matrix elements with $j_1=j_2$ are shown). The value of the matrix elements decreases rapidly with $n_2-n_1$. This facilitates restriction of the basis set by means of the principal quantum number.\label{fig:MatrixElements}}
\end{figure*}

\subsection{Matrix diagonalization}\label{sec:diagonalization}

In the previous sections, we established the Hamiltonian of two interacting Rydberg atoms and explained how matrix elements of the Hamiltonian can be calculated using \ref{sec:radial} and \ref{sec:angularmatrixelements}. This allows us to compute the matrix representation of the Hamiltonian. We calculate pair potentials of the unperturbed state  $\ket{n_1l_1j_1 m_{j1}; n_2l_2j_2 m_{j2}}$ by numerical diagonalization of the Hamiltonian matrix for a range of interatomic distances. Then, the eigenenergies are plotted versus the distances to make up the pair potentials. By drawing lines between the eigenenergy points for which the overlap between the eigenvectors is maximal, we extract the pair potential curves. The overlap between an eigenvector and the unperturbed state tells the probability to find the unperturbed state on the corresponding pair potential curve.

In order to make the diagonalization of the Hamiltonian computationally feasible, we have to keep the matrix size small. Therefore, it is important to exploit conservation laws. As discussed in section \ref{sec:conservations}, the Hamiltonian might conserve several quantum numbers and symmetries, leading to a block diagonal structure of the Hamiltonian matrix, see Figure \ref{fig:MatrixElements} \textsf{(a)}.

Each block can be diagonalized independently, which leads to a massive speed-up as the computation of all eigenpairs of a $n \times n$-matrix costs $\mathcal O(n^3)$ floating-point operations \cite{Bientinesi2009, Demmel2008}. Furthermore, we diagonalize only those blocks that belong to the unperturbed state we are interested in. In addition, we have to restrict the basis to states having a significant influence on the pair potentials of the unperturbed state.

Hereto, we have several possibilities. First, we can restrict the basis to elements with similar energies as the unperturbed state. Second, it is often useful to constrain the momentum quantum numbers because interaction of high order is required to change the momentum quantum numbers by large values. However, this constraint does not work if states involved in the high order interaction are degenerate as it is the case for Stark map calculations. Third, we can constrain the principal quantum number. If two states do not have similar principal quantum numbers, their radial matrix elements and hence their interaction is negligible, see Figure \ref{fig:MatrixElements} \textsf{(b)}. Higher order interactions, which would only require matrix elements with similar principal quantum numbers, would also be weak as the required order increases with the principal quantum number. How the constraints should be chosen depends on many factors like the considered distances, the required accuracy, the quantum numbers of the unperturbed states, the order of the multipole expansion, external fields, and the atomic species. Thus, giving the right constraints is difficult a priori. In fact, the constraints have to be tested for each calculation. For obtaining a small basis size, we start with strong constraints and loosen them systematically until the pair potentials of the unperturbed state have converged within the distance threshold we are interested in. The convergence is extensively discussed in \cite{Deiglmayr2016b}.

For calculating pair potentials in the presence of external electromagnetic fields, the following approach has turned out to be useful: We construct a single-atom basis and calculate the atom-field Hamiltonian for each of the two atoms independently. We enlarge the basis sets until the Stark/Zeeman maps have converged. We combine the eigenstates of the single-atom Hamiltonians into pair states and obtain a pair basis that is suitable for establishing the total Hamiltonian including atom-atom interactions. The combined eigenenergies, that are the Stark/Zeeman energies of the pair states, are located on the diagonal of the Hamiltonian. This procedure has the advantage, that we can now restrict the pair basis to the states relevant for the atom-atom interaction without losing accuracy in the treatment of the electromagnetic fields. Constraining the pair basis stronger than the single-atom basis is appropriate, in particular because the atom-field interaction typically couples over larger energy ranges than the atom-atom interaction. Furthermore, whereas atom-field interactions might change energies drastically, the small amount of admixed states is irrelevant for the atom-atom interaction in many cases of practical relevance. Despite these actions, the inclusion of electromagnetic fields can drastically increase the computational cost. Depending on the type of field, the block structure of the Hamiltonian matrix is destroyed.

\section{Applications of the Rydberg potential calculation}\label{sec:Applications}
In this section, we discuss two examples of the Rydberg potential calculation with relevance to recent experiments. Comparison with experimental results enables us to validate our numerical results and demonstrate the applicability of full potential calculations to state-of-the-art experiments. It also allows us to benchmark the influence of the basis size and of the truncation order on the interaction potentials.

\subsection{Relevance of higher-order multipole terms and basis size}
\begin{figure*}
	\begin{center}
		\includegraphics{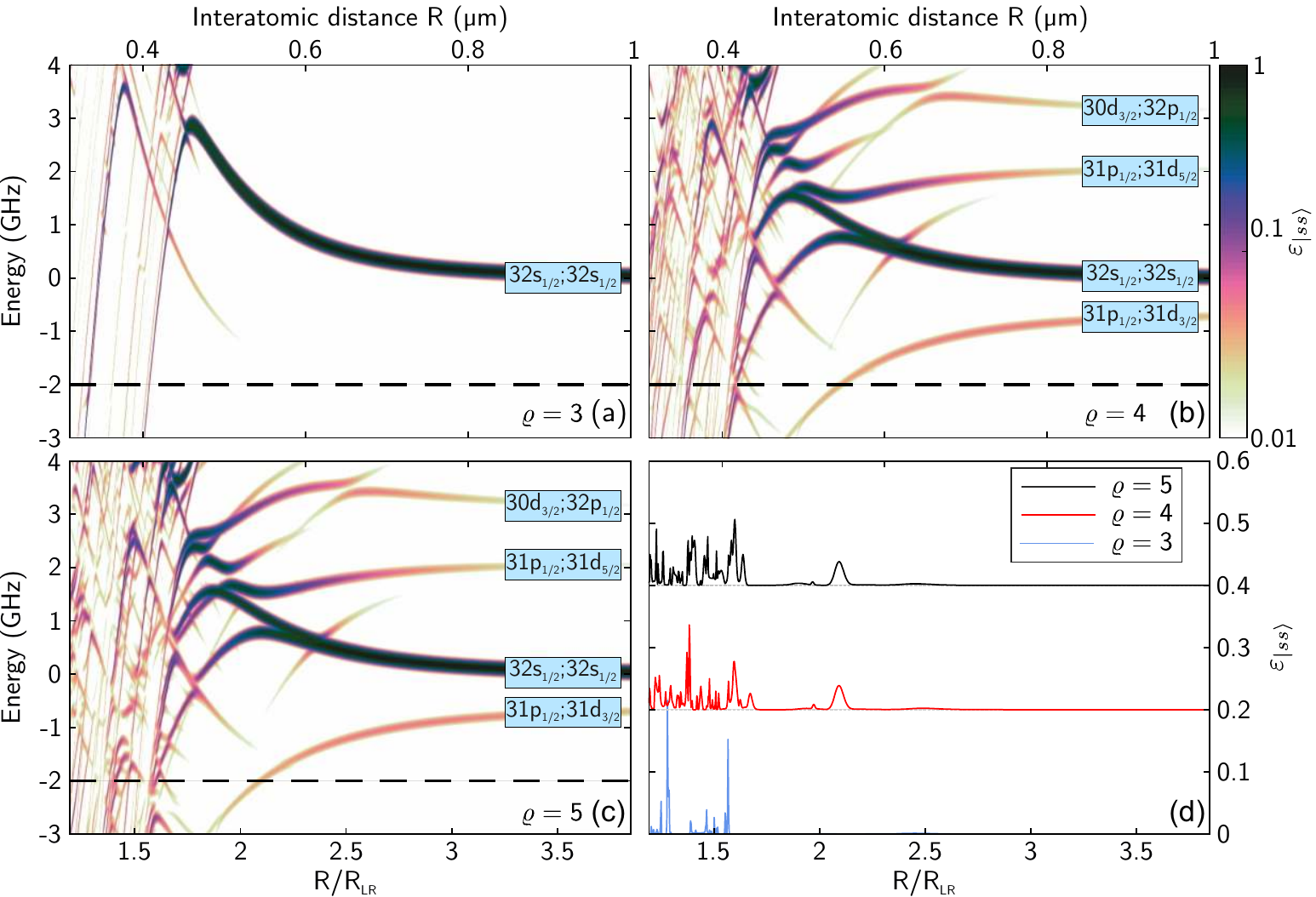}
		\caption{\label{fig:Cs32S} \textsf{(a-c)} Potential landscape around the unperturbed Cs $|32s_{1/2};32s_{1/2}\rangle$ state calculated up to order $1/R^\varrho$ of the multipole expansion of the interaction potential: \textsf{(a)} $\varrho=3$, \textsf{(b)} $\varrho=4$, \textsf{(c)} $\varrho=5$. \textsf{(d)} Admixture $\varepsilon_{|ss\rangle}$ to the perturbed pair states for a cut through the potential at a red detuning of $\SI{-2}{\giga\hertz}$. Cuts for $\varrho=4$ and $\varrho=5$ are shifted by an offset of $0.2$ and $0.4$, respectively. The inclusion of the dipole-quadrupole term ($\varrho=4$) results in the resonance feature at $R/R_\mathrm{LR} \approx 2.1$, which is identified in \cite{Loew2015} as the dominant underlying reason for the experimentally observed formation of Rydberg aggregates. While inclusion of one additional order significantly changes the potential landscape, this resonance feature is not affected.}
	\end{center}
\end{figure*}

\label{sec:higherOrders}
The relevance of multipole terms in the interaction potential of order higher than dipole-dipole, i.e. $\varrho >3$ in equation (\ref{eqn:multipoleorder}), has been highlighted in several recent experiments~\cite{Shaffer2006,Merkt2014,Loew2015,Deiglmayr2016b}. As an example, we focus here on the observation of Rydberg aggregation dynamics in a vapor cell at room temperature by Urvoy \textit{et al.} \cite{Loew2015}. In this experiment, the high atomic densities and the spectral width of the laser pulses allow to probe Rydberg interaction at short interatomic distances, with one key finding being that the correlated excitation of Rydberg atoms is driven by the dipole-quadrupole ($\varrho =4$) contribution to the interaction.

Specifically, in the experiment, Rydberg excitation in a Cesium vapor cell is driven by two-photon excitation with red detuning $\Delta=\omega_{\mathrm{Laser}} - \omega_{\mathrm{Atom}} = \SI{-2}{\giga\hertz}$ relative to the $32s$ Rydberg state. The pure van-der-Waals interaction potential resulting from dipole-dipole coupling of two atoms in this state is repulsive, see Figure~\ref{fig:Cs32S}~\textsf{(a)}, suggesting that the presence of one Rydberg atom does not increase the excitation probability of further Rydberg atoms by the red-detuned driving lasers.

However, including the dipole-quadrupole interaction ($\varrho =4$) results in admixture of the $|32s_{1/2};32s_{1/2}\rangle$ pair state into several close-lying, attractively interacting pair states (such as $|31p_{1/2};31d_{j}\rangle$ and $|32p_{1/2};30d_{j}\rangle$), as shown in Figure~\ref{fig:Cs32S}~\textsf{(b)}. As done in  \cite{Loew2015}, we quantify the admixture of, for example, $|ss\rangle = |32s_{1/2},m_j=1/2;32s_{1/2},m_j=-1/2\rangle$ to any Rydberg pair state $|\Psi\rangle$ by $\varepsilon_{|ss\rangle}(\Delta)= |\langle \Psi|ss\rangle|$. Any such admixture at detuning $\Delta$ results in efficient optical excitation of additional Rydberg atoms at specific distances to a first seed excitation. In particular, Urvoy \textit{et al.} identified the resulting resonance at $R/R_\mathrm{LR} \approx 2.1$ (where $R_\mathrm{LR}$ is the Le~Roy radius), which is reproduced by our calculations, as the dominant underlying mechanism for the correlated Rydberg aggregation observed in the experiment \cite{Loew2015}.

Based on this finding, an obvious question is how additional multipole orders further modify the interaction potential. In Figure~\ref{fig:Cs32S}~\textsf{(c)}, we show the resulting potential map when the $\varrho=5$ terms, corresponding to quadrupole-quadrupole and dipople-octupole interactions, are included. Significant effects of these contributions can be seen at small interatomic distances  $1<R/R_{\mathrm{LR}}<1.7$, for example in the detuning region between $\SI{2}{\giga\hertz}$ and $\SI{4}{\giga\hertz}$. For the experiment, the relevant figure is $\varepsilon_{|ss\rangle}(\Delta=\SI{-2}{\giga\hertz})$, we show the extracted values for all three potential calculations ($\varrho=3,4,5)$ in Figure~\ref{fig:Cs32S}~\textsf{(d)}.
While the inclusion of the $\varrho=5$ terms also modifies $\varepsilon_{|ss\rangle}$ at short distances, the main relevant resonance feature at $R/R_\mathrm{LR} \approx 2.1$ is not modified by the higher-order terms. Thus, the quantitative differences in the potential landscape due to the next higher-order terms do not affect the conclusions in~\cite{Loew2015}.

In contrast, the features at small distances are relevant for example for formation of bound pair states of Rydberg atoms~\cite{Gould2003,Shaffer2006,Gould2008,Shaffer2009,Deiglmayr2016}, requiring inclusion of even further orders in the calculation \cite{Deiglmayr2016,Deiglmayr2016b}. In general, when increasing $\varrho$, care has to be taken that the pair state basis truncation is appropriately adapted to include enough coupled states. In this example, we have used the constraints $\Delta n = 5$ and $\Delta l =6$ on the differences in quantum numbers of the individual Rydberg states with respect to the state $|32s_{1/2};32s_{1/2}\rangle$. These cutoff criteria are motivated by the selection rules for the different interaction orders discussed in section \ref{sec:conservations} and the scaling of the electric multipole matrix elements, see section \ref{sec:diagonalization}. Of particular importance for precise calculations at distances $1<R/R_{\mathrm{LR}}<1.7$ are the symmetry considerations of section \ref{sec:conservations}, which help to greatly reduce the size of the relevant pair state basis \cite{Deiglmayr2016b}.

\subsection{Angular dependence of the interaction near a F{\"o}rster resonance}
\begin{figure*}
	\begin{center}
		\includegraphics{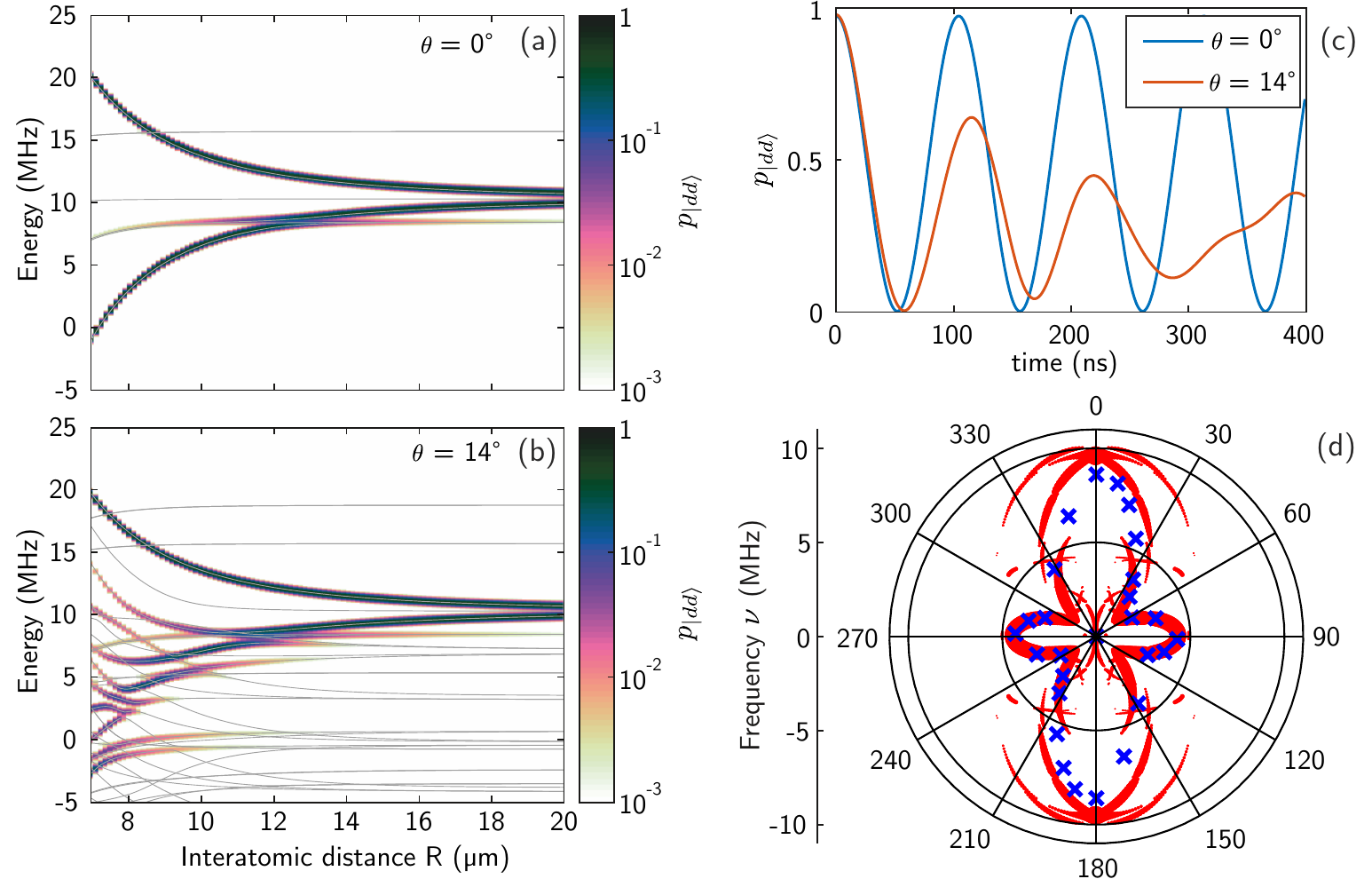}
		\caption{\label{fig:simulationAntoine} \textsf{(a)} Pair potential of the $|dd\rangle = |59d_{3/2},m_j=3/2;59d_{3/2},m_j=3/2\rangle$ state tuned into F\"{o}rster resonance with the $|pf\rangle = |61p_{1/2},m_j=1/2;57f_{5/2},m_j=5/2\rangle$ state by applying an electric field of $\SI{34.3}{\milli\volt/\centi\metre}$ for both atoms aligned along the quantization axis, i.e. $\theta = 0^\circ$. \textsf{(b)} Same pair potential as in \textsf{(a)} but for an angle of $\theta=14^\circ$ between the interatomic and the quantization axis.  \textsf{(c)} Time evolution of the probability to find the system in the $|dd\rangle$ state in the presence of an electric field. For $\theta = 0^\circ$ the system undergoes undamped oscillations between the $|dd\rangle$ and $|pf\rangle_{+/-}$ state with a frequency of $\SI{9.2}{\mega\hertz}$. The multi level structure relevant for $\theta = 14^\circ$ in \textsf{(b)} results in damping out of the oscillations due to dephasing (red line). \textsf{(d)} Angular dependence of the multiple oscillation frequencies out of the $|dd\rangle$ state. To illustrate how strong different frequencies do show up in the time evolution the size of the points encodes the relative weight of each frequency.}
	\end{center}
\end{figure*}

\label{section:angularexample} In the second example, we demonstrate the calculation of anisotropic Rydberg interactions in the presence of electric and magnetic fields, as discussed in sections \ref{sec:externalfields} and \ref{sec:angulardependency}. In this context, we calculate the interaction potentials measured in the experiments of Ravets \textit{et al.} in~\cite{Browaeys2015b}. Here, two single $^{87}Rb$ atoms were prepared in their ground state in two tightly focussed optical tweezers. Both the distance $R$ between the two atoms and the angle $\theta$ between the interatomic axis and the external fields could be precisely tuned. Using a two-photon excitation scheme, both atoms were excited to the $|59d_{3/2},m_j=3/2\rangle$ state by applying a $\pi$-pulse. In the pair state basis and at zero electric field, the state $|dd\rangle = |59d_{3/2},m_j=3/2;59d_{3/2},m_j=3/2\rangle$ is detuned by $\SI{8.69}{\mega\hertz}$ from the state $|pf\rangle = |61p_{1/2},m_j=1/2;57f_{5/2},m_j=5/2\rangle$. Due to the different polarizabilities of the states, both pair states could be tuned into degeneracy by applying a weak electric field of $\SI{34.3}{\milli\volt/\centi\meter}$. With this approach, Ravets \textit{et al.} could map out the angular shape of the electric dipole-dipole interaction between the two atoms \cite{Browaeys2015b}.

More specifically, the strength of the interaction was measured by letting the two-atom system evolve after the Rydberg excitation and in the presence of the electric field. After a variable hold time, a second optical $\pi$-pulse coupling to the $|dd\rangle$ state was employed to bring the atoms back to their ground state. By measuring the ground-state population after the full sequence, the time-evolution of the $|dd\rangle$ Rydberg pair state population could be reconstructed. Performing this experiment for various angles $\theta$ and fixed distance $R=\SI{9.1}{\micro\meter}$ resulted in the beautiful dipole-dipole pattern of the interaction shown by the blue crosses in Figure~\ref{fig:simulationAntoine} \textsf{(d)}.

For comparison with the experimental results, we calculate the full potentials for different angles $\theta$ including the finite electric and magnetic fields. Here, the optimized matrix construction discussed in section \ref{sec:diagonalization} is particularly relevant to make precise calculations feasible. As an example, we show in Figure~\ref{fig:simulationAntoine} \textsf{(a)} and \textsf{(b)} the potentials obtained for $\theta = 0^\circ$ (atoms aligned with respect to the external fields) and $\theta = 14^\circ$, respectively. Besides the energy shifts caused by the interaction, we encode here the probability $a_k = |\langle dd|\varphi_k\rangle|^2$ to find an admixture of the initially prepared unperturbed $|dd\rangle$ state in the new eigenstate $|\varphi_k\rangle$ as a density plot. For $\theta=0^\circ$, Figure~\ref{fig:simulationAntoine} \textsf{(a)} show that a two-level approximation is valid for most of the distances between $\SI{7}{\micro\meter}$ and $\SI{20}{\micro\meter}$. Most importantly, at the experimentally relevant distance $R=\SI{9.1}{\micro\meter}$, the system can be treated as a two-level system.

However, the situation changes for an angle of $\theta = 14^\circ$, where the two-level approximation breaks down. This is caused by mixing of the different fine structure states of the $|57f\rangle$ state and of different magnetic levels coupled for non-zero interaction angles.

From the calculated potentials and more specifically the overlap probabilities $a_k$, it is straight-forward to calculate the coherent evolution of the interacting Rydberg atom pair. Specifically the time-dependent probability of being in the $|dd\rangle$ state in the presence of interaction is given by $p_{|dd\rangle}(t) = |\sum_k a_k\exp\left(i \frac{E_k}{\hbar} t\right)]|^2$. Examples for two different angles and $R=\SI{9.1}{\micro\metre}$ are shown in Figure~\ref{fig:simulationAntoine}~\textsf{(c)}. For $\theta = 0^\circ$, we obtain an undamped sinusoidal oscillation with a frequency of $\nu = \SI{9.2}{\mega\hertz}$, which corresponds to the splitting of the most strongly populated pair potentials. In the case of $\theta = 14 ^\circ$ the significant coupling to multiple other pair states leads to dephasing that effectively damps out the Rabi oscillations. These results agree very well with the experimental time-evolution reported in~\cite{Browaeys2015b}.

In Figure~\ref{fig:simulationAntoine}~\textsf{(d)}, we show all frequencies contributing to the time evolution for $R=\SI{9.1}{\micro\metre}$ and varying $\theta$, obtained from the energy differences $E_m - E_n$ of the pair states to which the initial state couples (red points). The size of each point encodes the relative weight of each frequency, which is proportional to $a_m\cdot a_n$. For comparison, the single frequencies at each $\theta$ extracted from the experiment are shown by the blue crosses. One can see that for $0^\circ \leq \theta \leq 5^\circ$ and  for $55^\circ \leq \theta \leq 90^\circ$ our calculations find a single dominant contribution, which is in excellent agreement with the experimental data. In contrast, for angles outside these regions, the increased number of pair states contributing to the time evolution explains the damped oscillations measured in the experiment.

\section{Conclusions and outlook}
In this tutorial, we have reviewed the calculation of interaction potentials between pairs of atoms excited to Rydberg states. Precise knowledge of the full potential landscape has become relevant to a wide range of experiments utilizing Rydberg atoms in recent years. Our goal has been to give a comprehensive summary of all the relevant calculation steps of the Rydberg potential for describing current and future experiments. We reviewed the symmetry properties of the interaction Hamiltonian and the selection rules of the different multipole orders. These considerations are crucial for efficient calculation of the interaction potentials. Two further aspects important for experiments are the angular dependence of the interaction and the inclusion of external magnetic and electric fields of arbitrary direction. Particularly, electric fields offer the powerful ability to strongly modify the Rydberg interaction by tuning to F\"orster resonances of Rydberg pair states. To complete our tutorial, we summarize the calculation of Rydberg wave functions and multipole matrix elements in the appendix.

In parallel to this tutorial, we have released our calculation software which implements all the discussed features as an open source project, \mbox{\url{https://pairinteraction.github.io/}}. It is our hope that this code is useful to members of the Rydberg community, either for comparison to their own calculations or as a general tool to explore the rich physics of Rydberg interaction. Our software is built such that further extensions are straight-forward, with one obvious next step being the inclusion of more species available for calculation, such as the alkaline earth metals \cite{Jones2007,Potvliege2012} or Holmium \cite{Molmer2008b,Saffman2015c}. The goal of our open source approach is to stimulate active participation of other developers. With the rapid progress of both experiments and theory investigating interacting Rydberg systems, it seems likely that more features of the Rydberg potentials will be explored and exploited in the future.

\begin{acknowledgments}
We thank Charles Adams, Przemyslaw Bienias, Rick van Bijnen, Antoine Browaeys, Johannes Deiglmayr, Hannes Gorniaczyk, Christian Gross, Julius de Hond, Jan Kumlin, Thierry Lahaye, Igor Lesanovsky, Weibin Li, Robert L\"{o}w, Thomas Niederpr\"{u}m, Herwig Ott, Asaf Paris-Mandoki, Tilman Pfau, Thibault Peyronel, Pierre Pillet, Thomas Pohl, Jonathan Pritchard, Georg Raithel, James Shaffer, Nikola \v{S}ibal\'{i}c, Johannes Zeiher for important discussions and for testing our pair interaction software. We are very thankful to Antoine Browaeys and Thierry Lahaye for providing their data on the angular dependence of the F\"{o}rster resonance. This work is funded by the German Research Foundation through Emmy-Noether-grant HO 4787/1-1, GiRyd project HO 4787/1-3, and SFB/TRR21 and the Ministry of Science, Research and the Arts of Baden-W\"{u}rttemberg through RiSC grant 33-7533.-30-10/37/1 and the European Union H2020 FET Proactive project RySQ (grant N. 640378). O.F. acknowledges support from the Minerva Foundation.
\end{acknowledgments}

\appendix
\section*{Appendix}

\begin{table*}
	\caption{\label{tab:qdefects}References (with year) used for the quantum defects in this manuscript and in our software.}
		\begin{tabularx}{\textwidth}{p{8em} XXXXX}
			\toprule
			& Li                          & Na                         & K                          & Rb                          & Cs                        \\
			\midrule
			$n$S$_{1/2}$      & \cite{Haroche1986} (1986)   & \cite{Bergeman1992} (1992) & \cite{Pendrill1981} (1981) & \cite{Fortagh2011} (2011)   & \cite{Merkt2016} (2016)   \\
			$n$P$_{1/2}$      & \cite{Haroche1986} (1986)   & \cite{Podnos1995} (1995)   & \cite{Niemax1983} (1983)   & \cite{Gallagher2003} (2003) & \cite{Merkt2016} (2016)   \\
			$n$P$_{3/2}$      & \cite{Haroche1986} (1986)   & \cite{Podnos1995} (1995)   & \cite{Niemax1983} (1983)   & \cite{Gallagher2003} (2003) & \cite{Merkt2016} (2016)   \\
			$n$D$_{3/2}$      & \cite{Johansson1958} (1958) & \cite{Podnos1995} (1995)   & \cite{Pendrill1981} (1981) & \cite{Fortagh2011} (2011)   & \cite{Haroche1982} (1982) \\
			$n$D$_{5/2}$      & \cite{Johansson1958} (1958) & \cite{Podnos1995} (1995)   & \cite{Pendrill1981} (1981) & \cite{Fortagh2011} (2011)   & \cite{Merkt2016} (2016)   \\
			$n$F$_{5/2,7/2}$  & \cite{Johansson1958} (1958) & \cite{MacAdam1997} (1997)  & \cite{Risberg1956} (1956)  & \cite{Gallagher2006} (2006) & \cite{Sansonetti1987} (1987) \\
			$n$G$_{7/2,9/2}$  & -                           & \cite{MacAdam1997} (1997)  & -                          & \cite{Martin2006} (2006)    & \cite{Sansonetti1987} (1987) \\
			$n$H$_{9/2,11/2}$ & -                           & \cite{MacAdam1997} (1997)  & -                          & -                           & -                         \\
			\bottomrule
		\end{tabularx}
\end{table*}

In the appendix, we briefly review the required steps to assemble the Hamiltonian (\ref{eqn:hamiltonian}) for the interaction potential calculation. As discussed in section~\ref{sec:hamiltonian}, this requires the Rydberg level energies (\ref{eq:RydberglevelEnergy}), and the matrix elements of the single-atom electric multipole operators (\ref{eqn:sphericalmultipole}). Exact calculations of both the energy spectrum and electron wave functions are possible only for the hydrogen atom. We consider Rydberg atoms with a single electron in a highly excited state (principal quantum number $n \gg 1$), which behave very similar to hydrogen, since the Rydberg electron is effectively bound to a core with charge number $Z -(Z -1)=1$, consisting of the actual nucleus (with charge $Z$) and $Z-1$ inner electrons screening the core charge. This results in expressions for the Rydberg level series which are only slightly modified from the well-known hydrogen result and there exist relatively simple approaches to calculating single-electron wave functions that capture the physics of Rydberg atoms very well. The reduction to a single-electron problem is particularly justified for the alkali atoms, because of the closed-shell structure of the inner electrons. But for sufficiently large $n$ this treatment also works well for other atomic species, such as the noble gases \cite{Swainson1991}, the alkaline earth metals \cite{Potvliege2012}, and even the lanthanides~\cite{Saffman2015c}.

\section{Rydberg energy levels and wave functions}\label{sec:potential}
We consider Rydberg states including spin-orbit coupling, because the fine-structure of Rydberg states with low angular momentum $l$ is well resolved in current experiments on interacting Rydberg
atoms. The Rydberg levels are specified by the quantum numbers of the single Rydberg electron, namely $n$ (principal quantum number), $l$ (orbital angular momentum), $j = l\pm 1/2$ (total angular
momentum), and $m_j$ (magnetic quantum number). We neglect the hyperfine splitting caused by the coupling of the electron angular momentum $j$ to the nuclear spin $I$. Although experiments are now
resolving hyperfine levels of low-$l$ Rydberg states up to very high principal quantum numbers $n\approx90$~\cite{Merkt2013} and the hyperfine structure of Rydberg states can be a matter of
importance in some quantum information experiments \cite{Saffman2015d}, the typical splitting $\Delta_\mathrm{hfs} < 1 \SI{1}{\mega\hertz}$ for $n\geq40$~\cite{Gallagher2003,Spreeuw2013} makes the
hyperfine level structure (so far) irrelevant for interactions between Rydberg atoms.

\subsection{Quantum defects}

The energy of Rydberg levels of other species can be concisely written in analogy to the Rydberg expression for hydrogen as
\begin{equation}
  E_{nlj} = - \frac{h c R^*}{n^{*2}},
\end{equation}
Here, $n^*$ is an effective, non-integer principal quantum number, which contains the species-dependent deviation from hydrogen, while
\begin{equation}
  R^* = \frac{1}{1 + m_e/M_{\mathrm{atom}}} R_\infty
\end{equation}
is the modified Rydberg constant taking into account the species dependent mass $M_\mathrm{atom}$ of the atomic core. The effective quantum number is found to depend only weakly on $n$, but is mainly determined by the orbital angular momentum $l$. More specifically, the energies of the Rydberg series can be parameterized by introducing the quantum defects $n^* = n-\delta_{nlj}$, which in turn are written as a series expansion of the form
\begin{equation}
\label{eq:quantumdefect}
  \delta_{nlj} = \delta_0
  + \frac{\delta_2}{(n-\delta_0)^2}
  + \frac{\delta_4}{(n-\delta_0)^4}
  + \frac{\delta_6}{(n-\delta_0)^6}
  + \cdots \; .
\end{equation}
The coefficients in this polynomial expression are obtained from fits to experimentally measured transition energies for specific species.  The fine-structure splitting is usually included in the quantum defects, which results in them depending on the quantum number $j$. The quantum defects decrease rapidly with increasing orbital angular momentum $l$, since for high-$l$ states the influence of the non-hydrogenic core on the single Rydberg electron orbit becomes less relevant. Thus, quantum defects have been experimentally determined for Rydberg states with low orbital angular momentum $l$, with the most precise data being available for the Alkali atoms. Table~\ref{tab:qdefects} lists the references for the coefficients of the quantum defects for the Alkalis used for the calculations in this tutorial. These quantum defects are also implemented by default in our software, but can be replaced or extended by new values. The accuracy of the quantum defects used to compute the potential energies of Rydberg states is a key element to the precise determination of Rydberg interaction potentials. Note that the closed-form expression~\ref{eq:quantumdefect} is motivated by, but differs from the analytic result obtained in quantum-defect-theory \cite{Seaton1958,Seaton1983} by the fact that $\delta_0$ appears in the higher-order terms instead of $\delta_{nlj}$. This truncation ``spoils the theoretical significance'' of the quantum defects \cite{Swainson1991}, but is of course necessary when fitting experimental data and provides a simple and elegant expression for the energy of Rydberg levels.

Since Rydberg levels become more hydrogen-like with increasing $l$, we can use the analytic expression for hydrogen fine-structure energies for levels without an experimentally determined quantum defect. In addition, we include a correction term proportional to the core dipole polarizability $\alpha_d$ of the considered species:
\begin{align}
E_{nlj} =& - \frac{h c R^*}{n^2}\left(1+\frac{\alpha^2}{n(j+1/2)}+\frac{\alpha^2}{n^2}\right) \nonumber\\
&- \frac{e^2}{(4 \pi \varepsilon_0)^2 a_0^4}\frac{3 \alpha_d}{4 n^3 l^5}.
\label{eq:highLenergies}
\end{align}
The derivation of this formula and a detailed discussion of the core polarizability can be found in \cite{Gallagher1994}.

\subsection{Rydberg electron wave function}

Reducing the Rydberg atom to a single electron orbiting an extended core consisting of the atomic nucleus and the inner electrons, enables us to calculate effective single-electron Rydberg wave functions. Most importantly, the more complex structure of the effective core does not lift the spherical symmetry of the problem, thus the usual separation of variables for the Rydberg electron wave function into radial and angular part holds. The angular part is solved analytically and, when fine-structure is included, given by the spin spherical harmonics

\begin{align}
&Y_{j \pm \frac{1}{2},\frac{1}{2}, j, m_j}=\frac{1}{\sqrt{2\left(j\pm\frac{1}{2}\right)+1}} \nonumber\\
&\qquad \times
\left(\begin{array}{c}
\mp\sqrt{j\pm \frac{1}{2}\mp m_j+\frac{1}{2}} Y_{j\pm\frac{1}{2},m_j-\frac{1}{2}}\\
\sqrt{j\pm \frac{1}{2}\pm m_j+\frac{1}{2}}Y_{j\pm\frac{1}{2},m_j+\frac{1}{2}}\end{array}\right).
\end{align}

Based on this expression, the angular part of the electric multipole moments can be calculated analytically, including the usual multipole selection rules. This is discussed in detail in \ref{sec:angularmatrixelements}. Non-relativistic quantum defect theory provides analytical solutions for the radial part, known as \emph{Coulomb functions} \cite{Damgaard1949,Seaton1983}. The basic idea is to consider large distances $r$ from the nucleus, where the screening of the inner electrons results in an effective core charge $Z=1$. There, the radial Schr\"{o}dinger equation reduces to the well-known hydrogen case, except that the energy eigenvalues of the bound states are fixed via the experimentally determined quantum defects. As a consequence, the resulting solutions depend on the (non-integer) effective principal quantum number $n^*$:
\begin{align}
  \Psi^{\mathrm{rad}}_{n^*l}(r) =& \left(\frac{1}{a_0}\right)^{\!3/2} \frac{1}{\sqrt{(n^*)^2 \Gamma(n^*+l+1) \Gamma(n^*-l)}} \nonumber\\
  &\times
  W_{n^*,l+1/2}\biggl( \frac{2 r}{n^* a_0} \biggr).
  \label{eq:CoulombFunctions}
\end{align}

Here, $\Gamma(z)$ is the Gamma function, and $W_{k,m}(z)$ is the Whittaker function. The Coulomb functions are obtained by solving the hydrogen Coulomb radial equation where the energies corresponding to non-integer principal quantum numbers have already been inserted. These are approximate wave functions with the correct behavior for large $r$ and the right binding energy. For the calculation of transition matrix elements between Rydberg states, these are the important criteria. A relativistic generalization of the quantum-defect theory exists~\cite{Cheng1979}, but for the high-$n$ Rydberg states of interest here, the modification of the radial wave function due to the fine-structure correction turns out to be negligible.

\begin{figure*}
	\centering
	\includegraphics{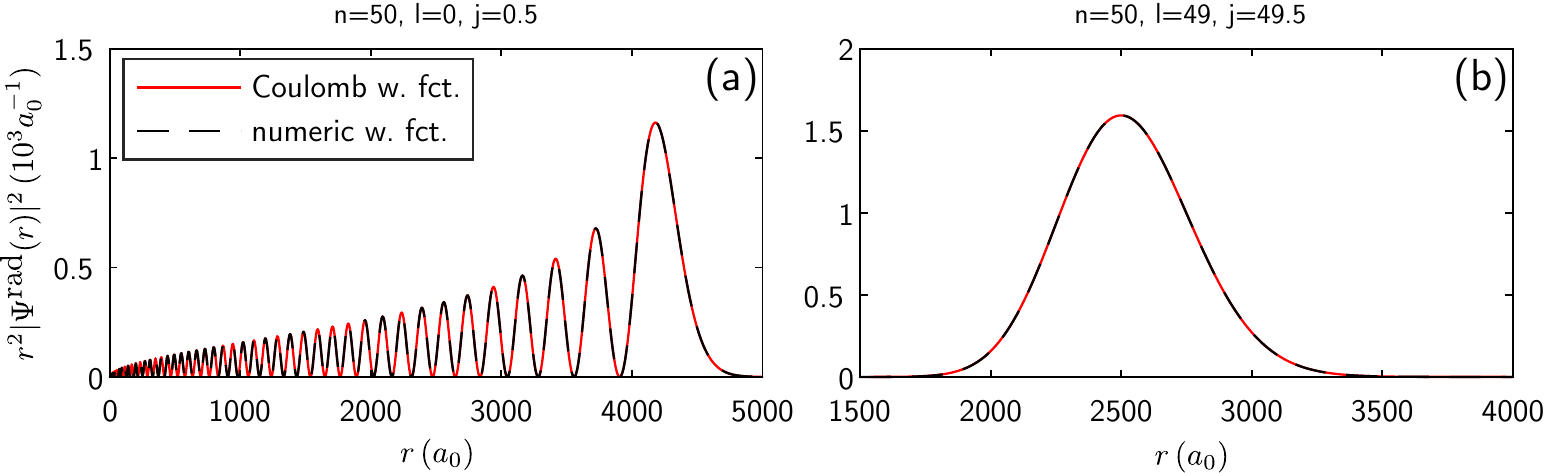}
	\caption{Comparison of radial Rydberg wave functions obtained via numerical integration of the Schr\"{o}dinger equation including the model potential (\ref{eq:Rydbergmodelpotential}) (dashed black line) and the corresponding Coulomb wave functions (\ref{eq:CoulombFunctions}) (red line). For the Rydberg states relevant for this tutorial, the Coulomb functions are highly accurate. For large $n$ though, the numerical wave functions can be calculated significantly faster.}
	\label{fig:comp-wf}
\end{figure*}

An alternative approach to obtaining single-electron wave functions is numerically solving the radial Schr\"{o}dinger equation including a species-dependent \emph{model potential}~\cite{Theodosiou1984}. Compared to quantum-defect theory this approach enables calculation of wave functions in the inner region if the model potentials were correctly determined. Typically, the model potential contains three contributions:
\begin{equation}
  V_{\mathrm{mod}}(r) = V_{\mathrm{C}}(r) + V_{\mathrm{P}}(r) + V_{\mathrm{s.o.}}(r).
  \label{eq:Rydbergmodelpotential}
\end{equation}
Here, $V_{\mathrm{C}}(r)$ is a modified Coulomb potential describing the distance dependent screening of the core charge by the inner electrons, $V_{\mathrm{P}}(r)$ describes the core polarization due to the Rydberg electron, and $V_{\mathrm{s.o.}}(r)$ is the spin-orbit coupling. The different terms are chosen such that the eigenvalues from the numerical solution of the radial Schr\"{o}dinger equation reproduce the experimentally measured Rydberg energies. If in turn the energies are fixed, the radial Schr\"{o}dinger equation reduces to a one-dimensional differential equation, and the electron wave functions can be obtained simply by numerical integration (usually from outside to inside). This approach, as well as the analytic Coulomb functions, are implemented in our software. In particular, for the alkali atoms we use expressions for $V_{\mathrm{C}}(r)$ and $V_{\mathrm{P}}(r)$ introduced by Marinescu \textit{et al.} \cite{Dalgarno1994}, which yield very good agreement with experimentally observed Rydberg level energies \cite{Sadeghpour2014}. In the model potential by Marinescu \textit{et al.}, the Coulomb interaction with the smeared out charge distribution of the inner shells is written as:
\begin{equation}
  V_{\mathrm{C}}(r) = - \frac{e^2}{4\pi\varepsilon_0} \frac{1 + (Z-1)\mathrm{e}^{-\alpha_1 r} - r(\alpha_3+\alpha_4 r)\mathrm{e}^{-\alpha_2 r}}{r},
\end{equation}
with coefficients $\alpha_{1,2,3,4}$ depending on the atomic species and the orbital angular momentum $l$~\cite{Dalgarno1994}. For the core polarization, only the leading dipole term is considered, which results in:
\begin{equation}
  V_{\mathrm{P}}(r) = -\frac{e^2}{(4\pi\varepsilon_0)^2} \frac{\alpha_d}{2 r^4} \left[ 1 - \mathrm{e}^{-(r/r_c)^6} \right].
\end{equation}
Here, $\alpha_d$ is again the core dipole polarizability and $r_c$ is the effective core size, obtained by comparing the numerical solutions with the experimentally observed energy levels. In addition to these two terms, we add an effective expression for the spin-orbit interaction~\cite{Greene1991}
\begin{equation}
  V_{\mathrm{s.o.}}(r > r_c) = \frac{1}{2} \left( \frac{e^2}{4\pi\varepsilon_0} \right) \left( \frac{g_s}{2 m_e^2 c^2} \right) \frac{\bm{l}\cdot\bm{s}}{r^3}.
\end{equation}
This expression is only valid for large $r$ and for smaller distances from the core the full expression derived from the Dirac equation has to be taken into account~\cite{Theodosiou1984}.

The spin-orbit interaction $V_{\mathrm{so}}$ depends on the radial coordinate $r$, thus the numerical radial wave function depends on the total angular momentum $j$. In practice, one usually does not solve the radial Schr\"odinger equation as eigenvalue problem, but instead inserts the level energies determined from experimental quantum defects (\ref{eq:highLenergies}). Here, care must be taken when combining model potentials (e.g. from \cite{Dalgarno1994}) with independently measured quantum defects, since the inserted energies most likely are not eigenenergies of the model potential. Improvements to the model potentials including the fine structure term have recently been discussed by Sanayei \textit{et al.}~\cite{Schopohl2015}.

We compare example wave functions obtained via numerical integration with the corresponding Coulomb wave functions in figure~\ref{fig:comp-wf}. The analytic Coulomb wave functions only indirectly include spin-orbit coupling and the modifications of the Coulomb potential from the simple hydrogen case via the quantum defects, while the model potential explicitly includes these effects in the Hamiltonian. For large $n$, and even more so for large $l$, the overlap of the Rydberg electron with the core region is vanishingly small, making the Coulomb functions very accurate solutions.

\begin{figure*}
	\centering
	\includegraphics{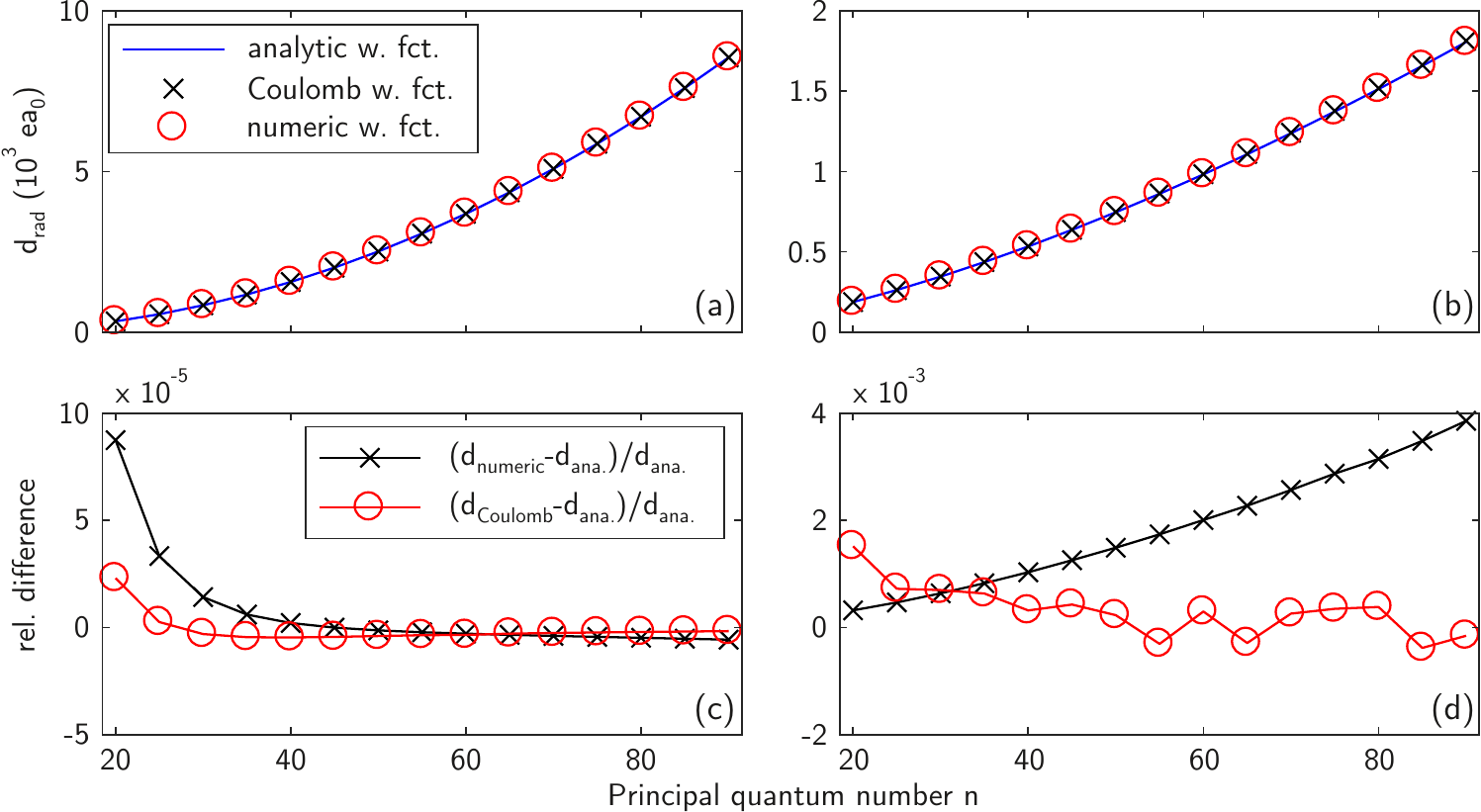}
	\caption{Comparison of radial dipole matrix elements $d_\textrm{rad} = \braket{nlj| \hat{p}^{\mathrm{rad}}_1 |n'l'j'}$ calculated by numeric integration using either Coulomb functions or model potential wave functions with the analytic expression from~\cite{Oumarou1998}. \textsf{(a+c)} show  dipole matrix elements for the $n, l=0,j=1/2 \leftrightarrow n'=n,l' = 1,j'=3/2$ transition and the relative difference between the three different approaches. All three methods are in very good agreement for the Rydberg states of interest here. \textsf{(b+d)} Dipole matrix elements for transitions from $n, l=n-1, j=l+1/2 \leftrightarrow n'=n, l'=n-2, j'=l'+1/2$ calculated for different principle quantum numbers and their relative difference. For transitions between high-$l$ states we observe a more significant systematic deviation between the result based on model potential wave functions and the other two approaches.}
	\label{fig:comp-dm}
\end{figure*}

\section{Radial matrix elements}\label{sec:radialmatrixelements}

\label{sec:radial} Calculating the radial parts of the electric multipole matrix elements appearing in the interaction Hamiltonian ((\ref{eqn:multipole}) in section~\ref{sec:hamiltonian}) amounts to solving integrals of the form
\begin{equation}
\braket{nlj| \hat{p}^{\mathrm{rad}}_\kappa |n'l'j'} = e \int \Psi^{\text{rad}}_{nlj}(r) \Psi^{\text{rad}}_{n'l'j'}(r) r^{2+\kappa} \; dr,
\end{equation}
where $\Psi^{\text{rad}}_{nlj}(r)$ are the radial wave functions discussed in~\ref{sec:potential}, obtained either numerically or in the form of Coulomb functions (\ref{eq:CoulombFunctions}), and $\kappa$ is the order of the multipole operator from (\ref{eqn:sphericalmultipole}) in section~\ref{sec:hamiltonian}. Note, that the radial wave functions obtained by either approach are real, so that the complex conjugation in the matrix element can be omitted. The matrix elements can be straightforwardly calculated by numerical integration~\cite{Damgaard1949,Kleppner1979}. To optimize the numerics it is useful to rescale the radial coordinate and the wave functions according to
\begin{equation}
x = \sqrt{r} \;,\quad X^{\text{rad}}_{nlj}(x) = x^{3/2} \Psi^{\text{rad}}_{nlj}(r).
\end{equation}
This scaling keeps the number of grid points between nodes of the wave function constant~\cite{Cooke1981}. As an alternative to numerical integration, various analytical expressions for electric dipole matrix elements exist~\cite{Cooke1981,Kaulakys1995,Oumarou1998}. In Figure~\ref{fig:comp-dm}, we compare electric dipole matrix elements obtained from numerical integration either using Coulomb functions or model potential wave functions and evaluation of the analytical expression in~\cite{Oumarou1998}. For transitions between low-$l$ states the three methods produce remarkable agreement for $n>40$. For high-$l$ transitions there are systematic deviations between the model potential results and the other two approaches, but the relative difference remains smaller than $1\%$.

It is important to note that the rather simple methods of calculating single-electron wave functions only yield accurate results for $n>30$. Significantly more advanced methods for calculating energy levels and matrix elements than what we present here have been developed for low-$n$ states, see e.g.~\cite{Safranova2011,Clark2016}.

\section{Angular matrix elements}
\label{sec:angularmatrixelements}
In addition to the radial part discussed in \ref{sec:radial}, we also need the angular part of the electric multipole matrix elements. In this appendix, we review the general formalism for calculating matrix elements of spherical tensor operators, which can be applied to determine the angular parts appearing when the multipole operators are expressed in the spherical basis. A more comprehensive discussion of this topic can be found for example in \cite{Sobelman1992}. The formalism relies on the Wigner-Eckart theorem \cite{Wigner1959}, which states that matrix elements of spherical tensor operators $\hat{T}_{\kappa q}$ can be expressed as products of a Wigner 3-j symbol (alternatively a Clebsch-Gordan coefficient) and a reduced matrix element, which is independent of the angular momentum orientation. As we perform calculations in the fine-structure basis, we show the Wigner-Eckart theorem for the total angular momentum $j = l + s$. It reads
\begin{align}
\langle l s j m_j | \hat{T}_{\kappa q} & | l' s' j' m_j' \rangle = (-1)^{j-m_j} (l s j||\hat{T}_{\kappa 0}||l' s' j') \nonumber\\
\times &  \left(\begin{array}{@{}lll@{}}
j & \kappa & j' \\
-m_j & q  & m_j' \\
\end{array}\right)\;, \label{eqn:WignerEckartCoupled}
\end{align}
where $(l s j||\hat{T}_{\kappa 0}||l' s' j')$ is the reduced matrix element for the total angular momentum. In case of the Wigner-Eckart theorem for the orbital angular momentum $l$ or spin $s$, the reduced matrix element $(l||\hat{T}_{\kappa 0}||l')$ or $(s||\hat{T}_{\kappa 0}||s')$ would occur instead.

If $\hat{T}_{\kappa q}$ commutes with the spin $s$, we can relate the different reduced matrix elements via the equation
\begin{align}
(l s j||\hat{T}_{\kappa 0} & ||l' s j')
=
(-1)^{l+s+j'+\kappa} (l||\hat{T}_{\kappa 0}  ||l') \nonumber\\
\times &  \sqrt{(2j+1)(2j'+1)}
\left\{\begin{array}{@{}lll@{}}
l & j & s \\
j' & l' & \kappa \\
\end{array}\right\} \;, \label{eqn:Reduced1}
\end{align}
where the last term is the Wigner 6-j symbol. If  $\hat{T}_{\kappa q}$ commutes with the orbital angular momentum $l$, we have the relation
\begin{align}
(l s j||\hat{T}_{\kappa 0} & ||l s' j')
=
(-1)^{l+s'+j+\kappa} (s||\hat{T}_{\kappa 0} ||s') \nonumber\\
\times &  \sqrt{(2j+1)(2j'+1)}
\left\{\begin{array}{@{}lll@{}}
s & j & l \\
j' & s' & \kappa \\
\end{array}\right\}\;. \label{eqn:Reduced2}
\end{align}

These equations facilitate the calculation of arbitrary matrix elements, provided that we know the value of the reduced matrix element $(l||\hat{T}_{\kappa 0}||l')$ or $(s||\hat{T}_{\kappa 0}||s')$, respectively. If the considered spherical tensor operator is a spherical harmonic $Y_{\kappa q}(\hat{\vartheta},\hat{\varphi})$, it commutes with the spin and  the value of the relevant reduced matrix element is
\begin{equation}
(l||\hat{Y}_{ \kappa 0}||l') =
(-1)^l \sqrt{\frac{(2l+1)(2\kappa+1)(2l'+1)}{4\pi}}
\left(\begin{array}{@{}lll@{}}
l & \kappa & l' \\
0 & 0 & 0 \\
\end{array}\right)\!.  \label{eqn:ReducedDipole}
\end{equation}
Given that spherical harmonics are proportional to the angular part of the multipole operator $\hat{p}_{\kappa q} ^\mathrm{ang} =  \sqrt{\frac{4 \pi}{2\kappa+1}} Y_{\kappa q}(\hat{\vartheta},\hat{\varphi})$, we can evaluate multipole matrix elements as well. In order to calculate matrix elements of the momentum operators $\hat{J}_{1q} \in \{\hat{l}_{1q},\hat{s}_{1q}\}$, we need the reduced matrix element
\begin{equation}
(J||\hat{J}_{10}||J') = \hbar \sqrt{J(J+1)(2J+1)}~\delta_{JJ'}\;. \label{eqn:ReducedMomentum}
\end{equation}

\end{document}